\documentclass[letterpaper,twocolumn,10pt]{article} %
\usepackage{amsmath}
\usepackage{usenix}
\newif\ifAnon\Anontrue
\newif\ifDraft\Draftfalse

\usepackage{todonotes}
\usepackage{cispa_defaults}
\usepackage{svg}
\usepackage{cpu_names}
\usepackage{stfloats}

\usepackage{subcaption}
\usepackage{multirow}
\usepackage{graphicx}
\usepackage{tcolorbox}
\usepackage{fontawesome5}
\usetikzlibrary{arrows.meta, positioning}
\usetikzlibrary{pgfplots.fillbetween} %
\usetikzlibrary{chains}

\newcommand{\code}[1]{\texttt{#1}\xspace}

\newcommand{\paragrabf}[1]{\noindent \textbf{#1.}\ }

\newcommand{\AttackName}{PhotoFinish\xspace}

\newcommand{\PlatformCount}{38\xspace}

\newcommand{\tprobe}{{\fontfamily{cmtt}\selectfont T0}\xspace}
\newcommand{\tref}{{\fontfamily{cmtt}\selectfont T1}\xspace}
\newcommand{\corS}{{\fontfamily{cmtt}\selectfont S}\xspace}
\newcommand{\corM}{{\fontfamily{cmtt}\selectfont M}\xspace}
\newcommand{\corO}{{\fontfamily{cmtt}\selectfont O}\xspace}
\newcommand{\rdelta}[1]{$\Delta_{#1}$}

\newlength{\maxheight}
\newcommand{\takeaway}[1]{%
\begin{tcolorbox}[boxsep=1pt,left=2pt,right=2pt,top=1pt,bottom=1pt]
    \faLightbulb[regular]~\textbf{Takeaway:} #1
\end{tcolorbox}
}

\begin{document}

\title{\AttackName: Portable Timerless Probing of Microarchitectural Latencies}

\author{
  {\rm Tristan Hornetz, Eduard Ebert, Lorenz Hetterich, Lukas Gerlach, Michael Schwarz}\\
  CISPA Helmholtz Center for Information Security
}

\maketitle

\begin{abstract}
Timing measurements are a fundamental building block of microarchitectural attacks, enabling software to convert otherwise invisible latency differences into architectural observations.
Thus, restricting high-resolution timers is a common defense, especially in constrained environments such as web browsers.
Timerless probing primitives provide an alternative approach.
Instead of reading or constructing a timer, they observe architectural events that correlate with the microarchitectural state.
However, such primitives typically rely on platform-specific behavior, and none generalizes across the wider computing landscape.

In this paper, we present \AttackName, a portable timerless probing primitive that converts latency differences directly into an architectural value.
\AttackName races an operation under test against a fixed reference in two parallel threads that then write unique values to the same shared location, resulting in a latency-dependent architectural value.
\AttackName only requires parallel execution and shared memory, without constructing or reading a clock.
We evaluate \AttackName on \PlatformCount CPUs spanning four ISAs and demonstrate reliable cache-state classification across widely different microarchitectures, with comparable fidelity when compiled to WebAssembly and executed inside a browser.
We further use \AttackName in AES T-table and \EvictReload attacks, show that it applies beyond caches to TLB and execution-engine latencies, and demonstrate sufficient precision to identify the LLC slice of an address.
Our results demonstrate shared-memory races as a generic and precise replacement for explicit timing, showing that restricting timers alone does not prevent software from observing fine-grained microarchitectural latency differences.
\end{abstract}

\section{Introduction}
Software-based microarchitectural attacks remain a practical threat to the confidentiality and integrity of modern systems.
They exploit effects below the architectural abstraction, such as cache state~\cite{Osvik2006,Yarom2014Flush}, branch-predictor state~\cite{Evtyushkin2018,Lee2017Inferring}, TLB state~\cite{Gras2018,Lipp2022Prefetch}, execution-unit contention~\cite{aldaya2019port,Rokicki2022port}, or DRAM row-buffer state~\cite{Pessl2016}.
Transient-execution attacks additionally exploit instructions that execute transiently and leave observable microarchitectural traces~\cite{Kocher2019,Lipp2018meltdown,Canella2019A}.
Although these effects are not directly visible in a program's architectural state, they affect execution time.
Thus, attackers can infer information across security boundaries by observing small latency differences in otherwise ordinary software execution.

A central building block of many such attacks is a timing measurement primitive.
The attacker executes an operation whose latency depends on microarchitectural state and then classifies it as fast or slow.
In cache attacks, this operation is commonly a memory access~\cite{Yarom2014Flush,Gruss2016Flush,Rauscher2025Systematic}.
In other attacks, it may be a branch~\cite{Aciicmez2007predicting,Lee2017Inferring,Evtyushkin2018}, an instruction sequence contending for a shared execution resource~\cite{aldaya2019port,Hornetz2025PortPrint}, a page-table walk~\cite{Gruss2016Prefetch,Lipp2022Prefetch}, or an instruction's execution time~\cite{Andrysco2015subnormal}.
In all cases, the attacker needs a way to turn a microarchitectural latency difference into an observation that software can use.

Traditionally, attackers obtain this observation using an incrementing timer.
The attacker timestamps the operation before and after execution and compares the difference against a fixed threshold.
Consequently, a common defense strategy is to restrict access to high-resolution timers: browsers reduce timer resolution and introduce jitter~\cite{Schwarz2017Timers,VanGoethem2015,Shusterman2021prime}, or disable them in specific settings~\cite{Intel2021sgxdeveloper,Lipp2016,thomas2026riscy}.
These restrictions do not remove the underlying latency difference.
They only remove the most direct way to observe it.
Attackers can compensate by amplifying the targeted difference~\cite{Allan2016,Roettger2021SpectreJS,Schwarzl2021dynamic,Purnal2023showtime} or constructing software clocks such as counter threads~\cite{Schwarz2017Timers,Lipp2016,Gras2017aslr}.

An alternative is to avoid constructing a timer altogether.
Existing timerless primitives expose microarchitectural state through specialized instructions~\cite{Disselkoen2017prime,kim2022dprime,Yu2023Synchronization,Zhang2023MWAIT,Thomas2025Exfilstate,Hornetz2026tdxray} or environment-specific properties of the cache hierarchy~\cite{Guanciale2016Cache,Jin2025Ghostcache,Thomas2025Exfilstate,thomas2026riscy}.
These techniques demonstrate that explicit timestamps are not essential, but they replace the dependence on a timer with assumptions about a particular ISA, processor feature, or cache implementation.
This leaves a portability gap: existing timerless primitives do not provide a generic way to observe latency differences across architectures and restricted execution environments, such as browsers.

In this paper, we ask the following research question:

\emph{Can we build a generic primitive that observes microarchitectural latency differences without reading a timer and without relying on specialized architecture-specific features?}

To answer this question, we present \AttackName, a timerless probing primitive applicable to all major computer architectures and even to restricted environments like web browsers.
We base this primitive on an observation by Purnal et al.~\cite{Purnal2023showtime}: A race condition between two parallel threads can reveal microarchitectural timing differences when both threads write to a shared variable after finishing a workload.
Specifically, one thread, which we call the \emph{probing thread} \tprobe, performs the target operation, whose latency depends on the microarchitectural state.
In parallel, a \emph{reference thread} \tref performs a reference operation with known latency.
Finally, both threads write distinct values to the same shared result variable.
When both threads complete, the final value of this variable reveals which store serialized last, showing whether the target operation was faster or slower than the reference.
The key property of this approach is that the attacker does not need a timestamp.
The primitive only requires two threads, shared memory, and sufficient control over instruction ordering.

With \AttackName, we develop this concept into a practical, generic, and near-universally applicable probing primitive for microarchitectural latencies.
We distinguish between a \emph{padded} and an \emph{unpadded} variant.
For sufficiently large latency differences, the unpadded variant works without prior calibration, which is typically required for timer-based primitives.
For smaller latency differences, padded \AttackName adds a calibrated delay to \tprobe.
This moves the race boundary in the same way that a threshold moves the decision boundary in a timer-based attack, but without ever reading a clock.

We evaluate \AttackName on \PlatformCount CPUs spanning x86\_64, AArch64, RISC-V, and LoongArch, and \num{45} platform configurations once hybrid architectures are separated.
This breadth allows us to demonstrate \AttackName's platform-agnostic properties:
For instance, unpadded \AttackName reliably separates L1 hits from DRAM accesses on most platforms, and padded \AttackName achieves this on all of them.
This is despite vastly different core microarchitectures, interconnect topologies, and even instruction sets.
A large set of test platforms also enables us to characterize \AttackName's performance broadly.
We identify the quantity that bounds the resolution of unpadded \AttackName -- the synchronization latency \rdelta{sync} between the two threads -- and measure it across our evaluation set.
This allows us to estimate the minimum latency difference for which unpadded \AttackName works.
By quantifying the jitter in \rdelta{sync} between runs, we can additionally estimate this for the padded variant.

The results show that \AttackName enables timerless variants, e.g., of \FlushReload~\cite{Yarom2014Flush} and \EvictReload~\cite{Lipp2016} on most platforms, a conclusion we also verify experimentally.
Padded \AttackName in particular features a high temporal accuracy, and can even distinguish adjacent cache levels on some target platforms.
When multiple measurements of a target latency are permitted, we show that \AttackName can reliably resolve timing differences at the level of individual clock cycles.
We demonstrate this by recovering an address's last-level-cache (LLC) slice on a \num{12}-slice Intel CPU with \SI{96.9}{\percent} accuracy.

\AttackName is not limited to native code.
We implement \AttackName in WebAssembly and evaluate it in Chromium for x86\_64, using only the standard \code{WebWorker} parallelism and \code{SharedArrayBuffer} memory.
We find that the synchronization latency strongly correlates with the native implementation (Pearson $r = \num{0.930}$ across \num{50} matched configurations), and directly verify that browser-based \AttackName can distinguish L1 hits from DRAM accesses on all tested machines.
This indicates that \AttackName's performance in the browser approaches that of the native implementation, making it a strong primitive for browser-based attacks.
Additionally, we show that \AttackName remains effective when implemented in Lua, achieving near-perfect classification of cached and uncached accesses despite JIT compilation.

We further compare \AttackName with counter-thread timers, a common fallback when explicit timers are unavailable.
Like counter threads, \AttackName remains affected by thread placement and cache-coherence behavior.
Thus, its advantage is not independence from these mechanisms, but that it directly encodes the microarchitectural state into an architectural value instead of constructing a clock from periodically visible counter updates, which requires a (potentially adaptive) threshold.
Across physical cores, this direct race usually provides a finer decision boundary and matches or slightly exceeds the classification accuracy of counting threads.
Padded \AttackName provides the broadest coverage, while SMT-sibling counting threads remain competitive for smaller LLC-to-DRAM latency differences.

Finally, we show that \AttackName substitutes for a timer in vastly different scenarios, including complete attacks.
We adapt a first-round AES T-table attack by replacing the timed reload step with \AttackName.
On an Apple M1, where unprivileged code sees only a \SI{24}{\mega\hertz} counter, we construct minimal L2 eviction sets and run an \EvictReload-based covert channel entirely without a high-resolution timer.
These case studies demonstrate that \AttackName is a practical primitive in environments where precise timers are restricted.
Furthermore, we show the primitive is not specific to caches by detecting TLB residency, AVX wakeup throttling, and subnormal floating-point penalties.

\textbf{Contributions.} Our contributions can be summarized as follows:
\begin{compactitem}
  \item We present \AttackName, a generic timerless primitive that converts microarchitectural latency differences into the architectural outcome of a shared-memory race, and implement it on x86\_64, AArch64, RISC-V, and LoongArch.

  \item We evaluate synchronization precision, latency classification, and counter-thread baselines on \PlatformCount systems across four ISAs, showing that \AttackName outperforms counting threads in many scenarios.

  \item We implement \AttackName in WebAssembly and demonstrate its efficacy inside the browser sandbox.

  \item We demonstrate that \AttackName can replace timed probing in practical settings, including an AES T-table attack and a timerless \EvictReload-style case study.
\end{compactitem}

\section{Background}\label{sec:background}
In this section, we provide the background required for the remainder of the paper.
We first introduce software-based cache side channels and the role of memory-access latency (\Cref{sec:background:cache-attacks}).
We then discuss timers and existing timer alternatives (\Cref{sec:background:timers}).
Finally, we briefly discuss shared-memory and cache-coherence behavior (\Cref{sec:background:coherence}).

\subsection{Cache Attacks}\label{sec:background:cache-attacks}
Modern CPUs use caches to compensate for high memory access latencies.
When a program accesses a memory address, the access is fast if the corresponding cache line is in a nearby cache, and slower if the line must be fetched from a slower cache level or main memory.
Although software does not see this directly, it affects execution time.
Cache attacks exploit this timing difference to infer information about another computation~\cite{Osvik2006,Yarom2014Flush,Gruss2016Flush,Rauscher2025Systematic}.
A typical cache attack consists of three phases.
First, the attacker prepares the cache state.
Second, the victim executes and may access memory depending on secret data.
Third, the attacker probes the cache state and infers whether the victim accessed a particular cache line.
The probing phase is usually implemented as a timing measurement: the attacker accesses an address and classifies the access as a cache hit or miss based on its latency. %
Thus, cache attacks observe secret-dependent cache state and translate it into information about the victim.

\paragrabf{\FlushReload}
\FlushReload relies on shared memory between attacker and victim~\cite{Yarom2014Flush}.
The attacker flushes a shared cache line from the cache hierarchy, waits for the victim, and reloads the line.
A fast reload indicates that the victim accessed the line in the meantime.
\FlushReload is precise because it monitors individual cache lines, but it requires either a flush instruction or another mechanism to evict the target line.

\paragrabf{\PrimeProbe and \EvictReload}
\PrimeProbe does not require shared memory.
Instead, the attacker fills a cache set with its own addresses, lets the victim execute, and then measures whether its own addresses were evicted~\cite{Osvik2006}.
\EvictReload combines the reload-based observation of \FlushReload with eviction sets instead of explicit flush instructions~\cite{Lipp2016}.
This makes it applicable on systems where flushing is unavailable or restricted, but it requires finding eviction sets that map to the same cache resources as the target address.

\subsection{Timers and Timer Alternatives}\label{sec:background:timers}
In the context of side-channel attacks, a timer is typically an incrementing counter that lets software compare the duration of two operations.
Native attackers can often use architectural timers, performance counters~\cite{Lipp2016}, or cycle counters~\cite{Gerlach2023A}, such as \code{rdtsc} on x86.
In constrained environments, such as browsers, precise timers are commonly restricted or coarsened because they enable microarchitectural side channels~\cite{Schwarz2017Timers,VanGoethem2015}.
Attackers use amplification and repeated measurement to address two distinct limitations of imprecise timers.
Amplification increases the latency difference until it exceeds the timer's quantization interval~\cite{Allan2016,Schwarzl2021dynamic,Purnal2023showtime,Katzman2023gates}.
Statistical aggregation instead reduces random measurement noise when the target permits repeated observations~\cite{Bernstein2005,Schwarz2019netspectre}.
Averaging cannot, by itself, recover a latency difference that deterministically maps to the same timer value.
It is only useful when noise or sampling jitter provides information across observations.
Both approaches increase the number of required operations or victim interactions, which is problematic when the interaction is short, rate-limited, or difficult to repeat.

\paragrabf{Counting Threads}
Counting threads~\cite{Schwarz2017Timers,Lipp2016,Gras2017aslr,Hetterich2022Branch} are a common fallback in microarchitectural attacks when hardware or the execution environment lacks a precise timer.
One thread continuously increments a shared variable, while the measuring thread reads this variable before and after the operation of interest.
The observed counter difference acts as a software clock.
Its effective resolution depends not only on the counter loop, but also on how quickly updates become visible to the measuring thread.
Consequently, thread placement, cache coherence, scheduler activity, contention, and CPU frequency changes can alter the relationship between counter increments and elapsed time.

\paragrabf{Event-Based Primitives}
Some prior attacks avoid conventional timers by observing other events.
For instance, weak coherence between data and instruction caches can reveal cache activity on some platforms~\cite{Guanciale2016Cache,Jin2025Ghostcache,Thomas2025Exfilstate,thomas2026riscy}.
On some architectures, this is also possible with specialized instructions such as the Intel TSX extensions~\cite{Disselkoen2017prime,kim2022dprime,Hornetz2026tdxray} or \texttt{umwait}~\cite{Zhang2023MWAIT}.
These approaches show that explicit timestamps are not strictly necessary for side-channel attacks, but typically rely on specific ISA features, processor extensions, or execution environments.

\subsection{Cache Coherence}\label{sec:background:coherence}

Multicore CPUs allow multiple cores to access the same memory location concurrently.
To keep memory coherent, CPUs implement a cache-coherence protocol that coordinates cache-line ownership~\cite{papamarcos1984low,hum2005forward,AMD64_vol2}.
A core that wants to write to a cache line typically has to obtain exclusive ownership of that line.
If the line is not already present in a writable state, the store can trigger coherence traffic, referred to as a read-for-ownership request~\cite{Intel_combined}.
Thus, even a cached store can have a variable latency depending on the initial cache state of the written cache line.
For a single memory location, coherence establishes an order in which writes become visible.
If two threads write different values to the same location, the final value observed in memory corresponds to the write that becomes visible last in this coherence order.
This final value is architecturally observable, although the exact write ordering may depend on microarchitectural effects such as cache state, interconnect latency, and core placement.
Modern CPUs may also reorder memory operations internally.
Out-of-order execution, store buffers, and weak memory models can cause loads and stores to be issued out of program order unless the implementation uses appropriate dependencies or fences~\cite{Sorin2011}.

\section{The \AttackName Primitive}\label{sec:primitive}

In this section, we introduce \AttackName in two variants.
We first present the unpadded primitive and derive its correctness condition (\Cref{subsec:primitive-overview}).
We then introduce a padded variant for smaller latency differences (\Cref{subsec:padded-probing}), discuss the sources of variability in the race (\Cref{subsec:sources-of-imprecision}), and describe our implementation and attacker requirements (\Cref{subsec:implementation,subsec:threat-model}).

\subsection{High-Level Design}\label{subsec:primitive-overview}
\AttackName uses two concurrent threads and a shared result word $w$.
The \emph{probing thread} \tprobe executes the target operation, whose latency is either $\Delta_{fast}$ or $\Delta_{slow}$.
The \emph{reference thread} \tref executes a fixed reference operation with latency $\Delta_{ref}$.
After their respective operations complete, both threads write distinct values to $w$.
Cache coherence serializes these writes, and the value that remains in $w$ identifies the store that became visible last.
By making each store data-dependent on the preceding operation, this architectural value reveals which execution path completed later.

\begin{figure*}[t]
  \centering
  \input{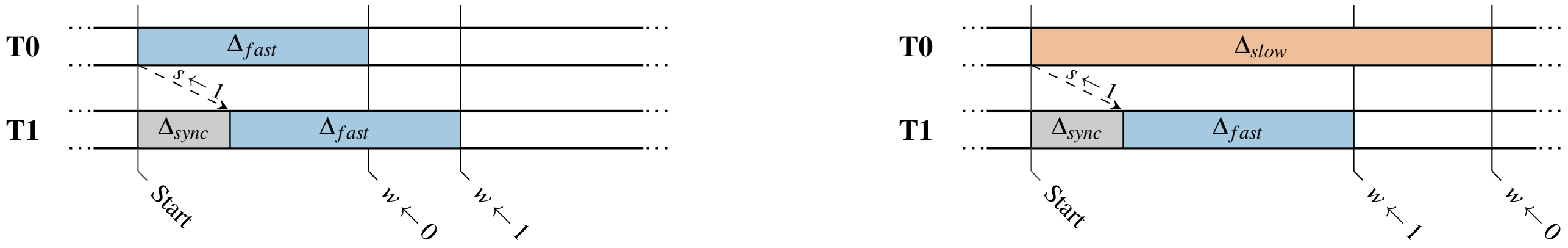}
  \begin{subfigure}{0.45\textwidth}
    \centering
    \begin{tikzpicture}
    \tikzset{arr/.style={->,>=stealth}}
    \tikzset{every node/.style={font={\footnotesize}}}
    \tikzset{placeholder/.style={inner sep=0,outer sep=0}}

    \node[placeholder] (l1l) at (0, 0) {};
    \node[placeholder] (l1r) at ($(l1l) + (\mainthreadwidth, 0)$) {};
    \node[placeholder] (l2l) at ($(l1l) + (0, \mainthreadheight)$) {};
    \node[placeholder] (l2r) at ($(l1l) + (\mainthreadwidth, \mainthreadheight)$) {};

    \node[placeholder] (l3l) at ($(l2l) + (0, \threaddistance)$) {};
    \node[placeholder] (l3r) at ($(l3l) + (\mainthreadwidth, 0)$) {};
    \node[placeholder] (l4l) at ($(l3l) + (0, \mainthreadheight)$) {};
    \node[placeholder] (l4r) at ($(l3l) + (\mainthreadwidth, \mainthreadheight)$) {};

    \node (lt0) at ($(l3l)!.5!(l4l) + (-\labeldist, 0)$) {\small\textbf{T0}};
    \node (lt1) at ($(l1l)!.5!(l2l) + (-\labeldist, 0)$) {\small\textbf{T1}};

    \draw[thick] (l1l) -- (l1r);
    \draw[dotted,thick] (l1l) -- ($(l1l) + (-\dotwidth, 0)$);
    \draw[dotted,thick] (l1r) -- ($(l1r) + (\dotwidth, 0)$);
    \draw[thick] (l2l) -- (l2r);
    \draw[dotted,thick] (l2l) -- ($(l2l) + (-\dotwidth, 0)$);
    \draw[dotted,thick] (l2r) -- ($(l2r) + (\dotwidth, 0)$);
    \draw[thick] (l3l) -- (l3r);
    \draw[dotted,thick] (l3l) -- ($(l3l) + (-\dotwidth, 0)$);
    \draw[dotted,thick] (l3r) -- ($(l3r) + (\dotwidth, 0)$);
    \draw[thick] (l4l) -- (l4r);
    \draw[dotted,thick] (l4l) -- ($(l4l) + (-\dotwidth, 0)$);
    \draw[dotted,thick] (l4r) -- ($(l4r) + (\dotwidth, 0)$);

    \draw [thin,color=black!70] ($(l4l) + (\startoffset, \lineoverhangtop)$) -- ($(l1l) + (\startoffset, -\lineoverhangbottom)$);
    \draw [thin,color=black!70] ($(l1l) + (\startoffset, -\lineoverhangbottom)$) -- ($(l1l) + (\startoffset + .1, -\lineoverhangbottom + -.1)$);
    \node[rotate=-45,anchor=west] (startlabel) at ($(l1l) + (\startoffset + .1, -\lineoverhangbottom + -.1)$) {Start};

    \draw [->,dashed,arr] ($(l3l) + (\startoffset, 0)$) -- node [sloped, anchor=south, inner sep=1pt, pos=.63]{\scriptsize $s \leftarrow 1$} ($(l2l) + (\syncoffset - .025, .025)$);
    \draw [fill=black!20] ($(l1l) + (\startoffset, 0)$) rectangle node [] {$\Delta_{sync}$} ($(l2l) + (\syncoffset, 0)$);

    \draw [fill=plotblue!40] ($(l3l) + (\startoffset, 0)$) rectangle node [] {$\Delta_{fast}$} ($(l4l) + (\startoffset +\lonewidth, 0)$);

    \draw [thin] ($(l4l) + (\startoffset + \lonewidth, \lineoverhangtop)$) -- ($(l1l) + (\startoffset + \lonewidth, -\lineoverhangbottom)$);
    \draw [thin] ($(l1l) + (\startoffset + \lonewidth, -\lineoverhangbottom)$) -- ($(l1l) + (\startoffset + \lonewidth + .1, -\lineoverhangbottom + -.1)$);
    \node[rotate=-45,anchor=west] (startlabel) at ($(l1l) + (\startoffset + \lonewidth + .1, -\lineoverhangbottom + -.1)$) {$w \leftarrow 0$};

    \draw [fill=plotblue!40] ($(l1l) + (\syncoffset, 0)$) rectangle node [] {$\Delta_{fast}$} ($(l2l) + (\syncoffset +\lonewidth, 0)$);

    \draw [thin] ($(l4l) + (\syncoffset + \lonewidth, \lineoverhangtop)$) -- ($(l1l) + (\syncoffset + \lonewidth, -\lineoverhangbottom)$);
    \draw [thin] ($(l1l) + (\syncoffset + \lonewidth, -\lineoverhangbottom)$) -- ($(l1l) + (\syncoffset + \lonewidth + .1, -\lineoverhangbottom + -.1)$);
    \node[rotate=-45,anchor=west] (startlabel) at ($(l1l) + (\syncoffset + \lonewidth + .1, -\lineoverhangbottom + -.1)$) {$w \leftarrow 1$};

\end{tikzpicture}
    \subcaption{Fast target: \tref writes last, yielding $w=1$.}
  \end{subfigure}
  \hfill
  \begin{subfigure}{0.45\textwidth}
    \centering
    \begin{tikzpicture}
    \tikzset{arr/.style={->,>=stealth}}
    \tikzset{every node/.style={font={\footnotesize}}}
    \tikzset{placeholder/.style={inner sep=0,outer sep=0}}

    \node[placeholder] (l1l) at (0, 0) {};
    \node[placeholder] (l1r) at ($(l1l) + (\mainthreadwidth, 0)$) {};
    \node[placeholder] (l2l) at ($(l1l) + (0, \mainthreadheight)$) {};
    \node[placeholder] (l2r) at ($(l1l) + (\mainthreadwidth, \mainthreadheight)$) {};

    \node[placeholder] (l3l) at ($(l2l) + (0, \threaddistance)$) {};
    \node[placeholder] (l3r) at ($(l3l) + (\mainthreadwidth, 0)$) {};
    \node[placeholder] (l4l) at ($(l3l) + (0, \mainthreadheight)$) {};
    \node[placeholder] (l4r) at ($(l3l) + (\mainthreadwidth, \mainthreadheight)$) {};

    \node (lt0) at ($(l3l)!.5!(l4l) + (-\labeldist, 0)$) {\small\textbf{T0}};
    \node (lt1) at ($(l1l)!.5!(l2l) + (-\labeldist, 0)$) {\small\textbf{T1}};

    \draw[thick] (l1l) -- (l1r);
    \draw[dotted,thick] (l1l) -- ($(l1l) + (-\dotwidth, 0)$);
    \draw[dotted,thick] (l1r) -- ($(l1r) + (\dotwidth, 0)$);
    \draw[thick] (l2l) -- (l2r);
    \draw[dotted,thick] (l2l) -- ($(l2l) + (-\dotwidth, 0)$);
    \draw[dotted,thick] (l2r) -- ($(l2r) + (\dotwidth, 0)$);
    \draw[thick] (l3l) -- (l3r);
    \draw[dotted,thick] (l3l) -- ($(l3l) + (-\dotwidth, 0)$);
    \draw[dotted,thick] (l3r) -- ($(l3r) + (\dotwidth, 0)$);
    \draw[thick] (l4l) -- (l4r);
    \draw[dotted,thick] (l4l) -- ($(l4l) + (-\dotwidth, 0)$);
    \draw[dotted,thick] (l4r) -- ($(l4r) + (\dotwidth, 0)$);

    \draw [thin,color=black!70] ($(l4l) + (\startoffset, \lineoverhangtop)$) -- ($(l1l) + (\startoffset, -\lineoverhangbottom)$);
    \draw [thin,color=black!70] ($(l1l) + (\startoffset, -\lineoverhangbottom)$) -- ($(l1l) + (\startoffset + .1, -\lineoverhangbottom + -.1)$);
    \node[rotate=-45,anchor=west] (startlabel) at ($(l1l) + (\startoffset + .1, -\lineoverhangbottom + -.1)$) {Start};

    \draw [->,dashed,arr] ($(l3l) + (\startoffset, 0)$) -- node [sloped, anchor=south, inner sep=1pt, pos=.63]{\scriptsize $s \leftarrow 1$} ($(l2l) + (\syncoffset - .025, .025)$);
    \draw [fill=black!20] ($(l1l) + (\startoffset, 0)$) rectangle node [] {$\Delta_{sync}$} ($(l2l) + (\syncoffset, 0)$);

    \draw [thin] ($(l4l) + (\startoffset + \dramwidth, \lineoverhangtop)$) -- ($(l1l) + (\startoffset + \dramwidth, -\lineoverhangbottom)$);
    \draw [thin] ($(l1l) + (\startoffset + \dramwidth, -\lineoverhangbottom)$) -- ($(l1l) + (\startoffset + \dramwidth + .1, -\lineoverhangbottom + -.1)$);
    \node[rotate=-45,anchor=west] (startlabel) at ($(l1l) + (\startoffset + \dramwidth + .1, -\lineoverhangbottom + -.1)$) {$w \leftarrow 0$};

    \draw [fill=plotblue!40] ($(l1l) + (\syncoffset, 0)$) rectangle node [] {$\Delta_{fast}$} ($(l2l) + (\syncoffset + \lonewidth, 0)$);

    \draw [thin] ($(l4l) + (\syncoffset + \lonewidth, \lineoverhangtop)$) -- ($(l1l) + (\syncoffset + \lonewidth, -\lineoverhangbottom)$);
    \draw [thin] ($(l1l) + (\syncoffset + \lonewidth, -\lineoverhangbottom)$) -- ($(l1l) + (\syncoffset + \lonewidth + .1, -\lineoverhangbottom + -.1)$);
    \node[rotate=-45,anchor=west] (startlabel) at ($(l1l) + (\syncoffset + \lonewidth + .1, -\lineoverhangbottom + -.1)$) {$w \leftarrow 1$};

    \draw [fill=plotred!40] ($(l3l) + (\startoffset, 0)$) rectangle node [] {$\Delta_{slow}$} ($(l4l) + (\startoffset + \dramwidth, 0)$);

\end{tikzpicture}
    \subcaption{Slow target: \tprobe writes last, yielding $w=0$.}
  \end{subfigure}
  \caption{Unpadded \AttackName. \tprobe releases \tref through the shared synchronization word $s$ and immediately executes the target operation. \tref observes the release after $\Delta_{sync}$ and executes the reference operation. In our memory-access instantiation, the target and reference addresses contain distinct marker values, $0$ and $1$, which are propagated to the final stores. The value remaining in $w$ therefore identifies the path that completed later.}
  \label{fig:primitive}
\end{figure*}

\paragrabf{Race Construction}
\Cref{fig:primitive} illustrates the race.
Initially, \tref spins on a shared synchronization word $s$.
\tprobe starts a measurement by writing to $s$ and immediately executing the target operation with latency $\Delta_x$.
\tref observes the write after the synchronization latency $\Delta_{sync}$ and then executes its reference operation with $\Delta_{ref}$.
Ignoring jitter and assuming marker values `0' and `1', the architectural result is
$$
  w =
  \begin{cases}
    1, &\Delta_x < \Delta_{sync} + \Delta_{ref},\\
    0, &\Delta_x > \Delta_{sync} + \Delta_{ref}.
  \end{cases}
  \label{eq:photofinish-outcome}
$$
In practice, the outcome of measurements close to equality can be influenced by system-noise-induced jitter in $\Delta_{sync}$, as evaluated in \Cref{subsec:delta-sync}.

\paragrabf{Correctness Condition}
The race distinguishes the fast and slow states whenever the reference path finishes between them, i.e., $\Delta_{fast} < \Delta_{sync} + \Delta_{ref} < \Delta_{slow}$.
Unless otherwise stated, we always choose the reference operation to match the known fast case, i.e., $\Delta_{ref}=\Delta_{fast}$, as illustrated in \Cref{fig:primitive}.
The left inequality holds by construction, and the remaining requirement is $\Delta_{slow} - \Delta_{fast} > \Delta_{sync}$.
Hence, $\Delta_{sync}$ determines the temporal resolution of unpadded \AttackName: smaller synchronization latency permits the primitive to separate smaller latency differences.
Inducing a reference delay of $\Delta_{fast}$ is often trivial in practice.
With memory-access probing, for instance, both the target's fast state and the reference address can be prepared as L1 hits.
Unlike a timer-based classifier, unpadded \AttackName does not require a platform-specific numerical threshold.
The reference operation provides the comparison point directly, and simultaneous execution makes the result a relative ordering rather than an elapsed-time measurement.
We show in \Cref{sec:evaluation} that the latency difference between an L1 hit and a DRAM access exceeds $\Delta_{sync}$ on almost all tested platforms.
While we focus primarily on memory access timings, \AttackName also applies to other microarchitectural latencies (cf. \Cref{subsec:other-latency-types}), and even architectural latencies like a function's execution time.

\takeaway{Unpadded \AttackName classifies a latency without a timer or numerical threshold whenever $\Delta_{slow}-\Delta_{fast}>\Delta_{sync}$.}

\subsection{Padded \AttackName}\label{subsec:padded-probing}

\begin{figure}[t]
  \definecolor{plotblue}{RGB}{31,119,180}
\definecolor{plotred}{RGB}{213,94,0}

\newcommand{\mainthreadwidth}{5}
\newcommand{\mainthreadheight}{.4}
\newcommand{\threaddistance}{.1}
\newcommand{\threadgroupdistance}{.3}
\newcommand{\dotwidth}{.25}
\newcommand{\labeldist}{2.25}

\newcommand{\startoffset}{.5}
\newcommand{\syncoffset}{1.5}
\newcommand{\lonewidth}{3}
\newcommand{\dramwidth}{3.25}

\newcommand{\padwidth}{\syncoffset - \startoffset - .125}

\newcommand{\lineoverhangtop}{.25}
\newcommand{\lineoverhangbottom}{.25}
  \centering
  \begin{tikzpicture}
    \tikzset{arr/.style={->,>=stealth}}
    \tikzset{every node/.style={font={\footnotesize}}}
    \tikzset{placeholder/.style={inner sep=0,outer sep=0}}

    \node[placeholder] (l1l) at (0, 0) {};
    \node[placeholder] (l1r) at ($(l1l) + (\mainthreadwidth, 0)$) {};
    \node[placeholder] (l2l) at ($(l1l) + (0, \mainthreadheight)$) {};
    \node[placeholder] (l2r) at ($(l1l) + (\mainthreadwidth, \mainthreadheight)$) {};

    \node[placeholder] (l3l) at ($(l2l) + (0, \threaddistance)$) {};
    \node[placeholder] (l3r) at ($(l3l) + (\mainthreadwidth, 0)$) {};
    \node[placeholder] (l4l) at ($(l3l) + (0, \mainthreadheight)$) {};
    \node[placeholder] (l4r) at ($(l3l) + (\mainthreadwidth, \mainthreadheight)$) {};

    \node[placeholder] (l5l) at ($(l4l) + (0, \threadgroupdistance)$) {};
    \node[placeholder] (l5r) at ($(l5l) + (\mainthreadwidth, 0)$) {};
    \node[placeholder] (l6l) at ($(l5l) + (0, \mainthreadheight)$) {};
    \node[placeholder] (l6r) at ($(l5l) + (\mainthreadwidth, \mainthreadheight)$) {};

    \node[placeholder] (l7l) at ($(l6l) + (0, \threaddistance)$) {};
    \node[placeholder] (l7r) at ($(l7l) + (\mainthreadwidth, 0)$) {};
    \node[placeholder] (l8l) at ($(l7l) + (0, \mainthreadheight)$) {};
    \node[placeholder] (l8r) at ($(l7l) + (\mainthreadwidth, \mainthreadheight)$) {};

    \node[placeholder] (l0r) at ($(l1r) + (0, -\threadgroupdistance - .15)$) {};
    \node[placeholder] (l0l) at ($(l0r) + (-\mainthreadwidth, 0)$) {};
    \node[placeholder] (lnr) at ($(l0r) + (0, -\mainthreadheight)$) {};
    \node[placeholder] (lnl) at ($(l0r) + (-\mainthreadwidth, -\mainthreadheight)$) {};

    \node[anchor=west] (lt0np) at ($(l6l)!.5!(l7l) + (-\labeldist, 0)$) {\small\textbf{T0}};
    \node[anchor=west] (lt0np) at ($(l2l)!.5!(l3l) + (-\labeldist, 0)$) {\small\textbf{T0 (padded)}};
    \node[anchor=west] (lt1) at ($(lnl)!.5!(l0l) + (-\labeldist, 0)$) {\small\textbf{T1}};

    \draw[thick] (l0l) -- (l0r);
    \draw[dotted,thick] (l0l) -- ($(l0l) + (-\dotwidth, 0)$);
    \draw[dotted,thick] (l0r) -- ($(l0r) + (\dotwidth, 0)$);
    \draw[thick] (lnl) -- (lnr);
    \draw[dotted,thick] (lnl) -- ($(lnl) + (-\dotwidth, 0)$);
    \draw[dotted,thick] (lnr) -- ($(lnr) + (\dotwidth, 0)$);

    \draw[thick] (l1l) -- (l1r);
    \draw[dotted,thick] (l1l) -- ($(l1l) + (-\dotwidth, 0)$);
    \draw[dotted,thick] (l1r) -- ($(l1r) + (\dotwidth, 0)$);
    \draw[thick] (l2l) -- (l2r);
    \draw[dotted,thick] (l2l) -- ($(l2l) + (-\dotwidth, 0)$);
    \draw[dotted,thick] (l2r) -- ($(l2r) + (\dotwidth, 0)$);
    \draw[thick] (l3l) -- (l3r);
    \draw[dotted,thick] (l3l) -- ($(l3l) + (-\dotwidth, 0)$);
    \draw[dotted,thick] (l3r) -- ($(l3r) + (\dotwidth, 0)$);
    \draw[thick] (l4l) -- (l4r);
    \draw[dotted,thick] (l4l) -- ($(l4l) + (-\dotwidth, 0)$);
    \draw[dotted,thick] (l4r) -- ($(l4r) + (\dotwidth, 0)$);

    \draw[thick] (l5l) -- (l5r);
    \draw[dotted,thick] (l5l) -- ($(l5l) + (-\dotwidth, 0)$);
    \draw[dotted,thick] (l5r) -- ($(l5r) + (\dotwidth, 0)$);
    \draw[thick] (l6l) -- (l6r);
    \draw[dotted,thick] (l6l) -- ($(l6l) + (-\dotwidth, 0)$);
    \draw[dotted,thick] (l6r) -- ($(l6r) + (\dotwidth, 0)$);

    \draw[thick] (l7l) -- (l7r);
    \draw[dotted,thick] (l7l) -- ($(l7l) + (-\dotwidth, 0)$);
    \draw[dotted,thick] (l7r) -- ($(l7r) + (\dotwidth, 0)$);
    \draw[thick] (l8l) -- (l8r);
    \draw[dotted,thick] (l8l) -- ($(l8l) + (-\dotwidth, 0)$);
    \draw[dotted,thick] (l8r) -- ($(l8r) + (\dotwidth, 0)$);

    \draw [thin,color=black!70] ($(l8l) + (\startoffset, \lineoverhangtop)$) -- ($(lnl) + (\startoffset, -\lineoverhangbottom)$);

    \draw [fill=black!20] ($(lnl) + (\startoffset, 0)$) rectangle node [] {$\Delta_{sync}$} ($(l0l) + (\syncoffset, 0)$);

    \draw [fill=plotblue!40] ($(l7l) + (\startoffset, 0)$) rectangle node [] {$\Delta_{fast}$} ($(l8l) + (\startoffset +\lonewidth, 0)$);
    \draw [fill=plotred!40] ($(l5l) + (\startoffset, 0)$) rectangle node [] {$\Delta_{slow}$} ($(l6l) + (\startoffset + \dramwidth, 0)$);

    \draw [fill=green!40] ($(l3l) + (\startoffset, 0)$) rectangle node [] {$\Delta_{pad}$} ($(l4l) + (\startoffset +\padwidth, 0)$);
    \draw [fill=green!40] ($(l1l) + (\startoffset, 0)$) rectangle node [] {$\Delta_{pad}$} ($(l2l) + (\startoffset + \padwidth, 0)$);
    \draw [fill=plotblue!40] ($(l3l) + (\startoffset + \padwidth, 0)$) rectangle node [] {$\Delta_{fast}$} ($(l4l) + (\startoffset +\padwidth+\lonewidth, 0)$);
    \draw [fill=plotred!40] ($(l1l) + (\startoffset + \padwidth, 0)$) rectangle node [] {$\Delta_{slow}$} ($(l2l) + (\startoffset + \padwidth +\dramwidth, 0)$);

    \draw [fill=plotblue!40] ($(lnl) + (\syncoffset, 0)$) rectangle node [] {$\Delta_{fast}$} ($(l0l) + (\syncoffset +\lonewidth, 0)$);

    \draw [thin] ($(l8l) + (\syncoffset + \lonewidth, \lineoverhangtop)$) -- ($(lnl) + (\syncoffset + \lonewidth, -\lineoverhangbottom)$);

\end{tikzpicture}
  \caption{Unpadded and padded \AttackName. If the target latency difference is smaller than $\Delta_{sync}$, both target states finish before the reference path. Adding a controlled delay $\Delta_{pad}$ to \tprobe shifts the race boundary so that only the fast target finishes first.}
  \label{fig:primitive_padded}
\end{figure}
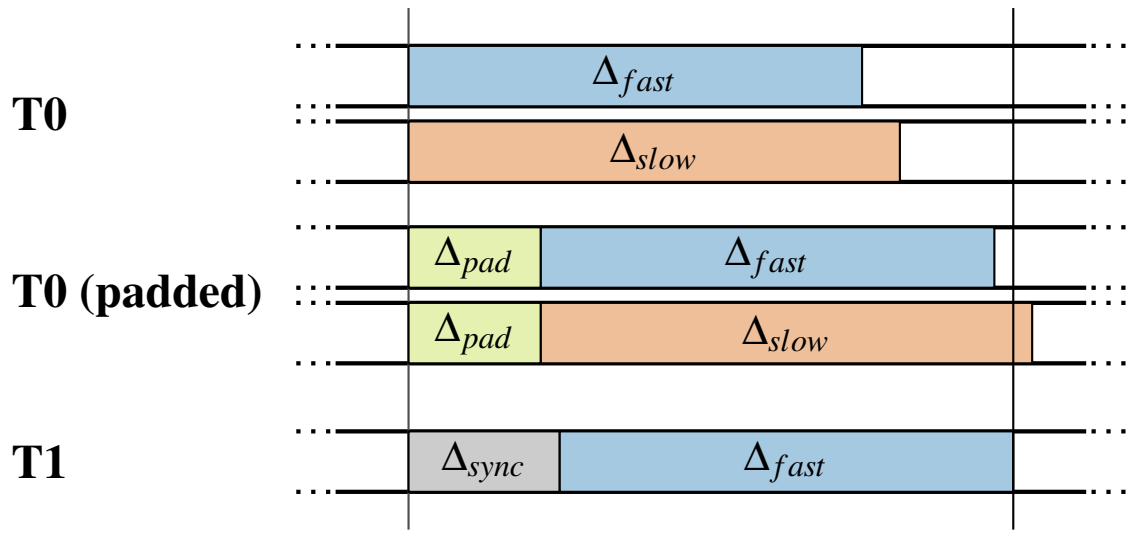

Unpadded \AttackName cannot separate two target states when both finish before the reference path.
This occurs, for example, when distinguishing adjacent cache levels whose latency difference is smaller than $\Delta_{sync}$.
Padded \AttackName addresses this limitation by adding a controlled delay $\Delta_{pad}$ to \tprobe before its final store, as illustrated in \Cref{fig:primitive_padded}.
A padding value is valid if $\Delta_{pad}+\Delta_{fast} < \Delta_{sync}+\Delta_{ref} < \Delta_{pad}+\Delta_{slow}$.
Assuming no jitter, such a value exists for every non-zero latency difference.
Thus, padding moves the decision boundary like a numerical threshold in a conventional timer, but without constructing or reading a clock.
We calibrate $\Delta_{pad}$ using known instances of the fast and slow states.
Starting from zero, we increase the padding until \tprobe wins the fast case while \tref wins the slow case.
A binary search can locate this interval efficiently: if \tref wins both cases, the padding is too large; if \tprobe wins both, it is too small.
The calibration depends on the platform and the target latency pair, but not on an external timing source.
We place the padding in \tprobe rather than replacing the reference operation with a calibrated delay.
A reference path cannot complete earlier than $\Delta_{sync}$ after the release.
Thus, a delay-only reference cannot distinguish two target latencies that are both shorter than $\Delta_{sync}$.
Padding \tprobe shifts the effective boundary below $\Delta_{sync}$ to cover this case as well.
In practice, the minimum distinguishable difference is bounded by jitter in $\Delta_{sync}$, $\Delta_{fast}$, and $\Delta_{slow}$.
However, only synchronization jitter is specific to \AttackName.
Variability in target latencies also affects traditional threshold-based timing classifiers.
We quantify the synchronization jitter and the resulting uncertain range in \Cref{subsec:delta-sync}.

\takeaway{Padded \AttackName relies on one-time calibration to separate latency differences smaller than $\Delta_{sync}$.}

\subsection{Controlling Jitter}\label{subsec:sources-of-imprecision}

There are three sources of timing jitter in \AttackName: the target and reference operations, the synchronization from \tprobe to \tref, and the coherence arbitration of the final writes to $w$.
The first is inherent to the probed event and also affects timer-based classifiers.
The latter two are specific to the race and can be reduced through careful state preparation and instruction ordering.

\paragrabf{Memory and Coherence State}
We place the synchronization word $s$ and result word $w$ on separate cache lines.
With that, releasing \tref through $s$ does not share a coherence transaction with the final write race on $w$.
Before each measurement, both threads touch the two lines, placing them in their private caches and normalizing the access path.
We additionally establish the same clean Shared (\corS) state on both cores, ensuring that either final writer must obtain ownership before transitioning the result line to Modified (\corM).
This preparation requires additional care on protocols with owned dirty sharing, such as MOESI~\cite{AMD64_vol2}.
If one core holds a line in Owned (\corO) while the other holds it in Shared (\corS), the owner may upgrade to Modified with less coherence traffic than the sharer, biasing the race.
Thus, we flush or evict $s$ and $w$ to the point of coherence before both threads reload them, ensuring that neither core retains ownership.
This step is unnecessary for MESI~\cite{papamarcos1984low} and MESIF~\cite{hum2005forward}, where loading the line on both cores already establishes a clean, non-owned shared state.

\paragrabf{Instruction Ordering}
We implement the latency-critical paths in assembly and connect the relevant operations through true data dependencies.
Otherwise, compiler transformations and out-of-order execution can move work across the race boundaries.
For memory-access probing, the value loaded from the target or reference address is itself propagated to the final store to $w$.
For more general operations, we carry a marker value through a dependency chain.
Note that integrating store operations into such a dependency chain is typically not possible, as they usually do not have an output register.
Hence, we implement stores with an exchange instruction if available on the target instruction set.

\paragrabf{Thread Placement and System Activity}
$\Delta_{sync}$ depends on the placement of \tprobe and \tref, including whether they share a physical core, cache hierarchy, interconnect path, or core complex.
Hence, scheduler migration would change the decision boundary during an experiment.
Our native implementation pins both threads to fixed logical cores, and we evaluate the effect of different placements explicitly in \Cref{subsec:delta-sync}.

\subsection{Implementation}\label{subsec:implementation}

\begin{figure}[t]
  \begin{minipage}[t]{0.48\linewidth}
    \centering
    \small\textbf{\tprobe: Probing Thread}
    \begin{lstlisting}[language={},numbers=none,frame=single,
      basicstyle=\scriptsize\ttfamily,
      commentstyle=\scriptsize\itshape\color{gray!60!black},
      morekeywords={await,while},
      keywordstyle=\bfseries\color{blue!60!black},
      columns=fullflexible,breaklines=false,
      xleftmargin=12pt,xrightmargin=4pt,
      framexleftmargin=0pt,framexrightmargin=0pt,
      numbers=left,numbersep=5pt,
      numberstyle=\tiny\color{gray}]
prepare_shared(s, w)

# pre-synchronize
pre_s = 1
await pre_s == 0

# execute target
s = 1

r = target_op()   # marker 0
w = r
    \end{lstlisting}
  \end{minipage}
  \hspace{1em}
  \begin{minipage}[t]{0.43\linewidth}
    \centering
    \small\textbf{\tref: Reference Thread}
    \begin{lstlisting}[language={},numbers=none,frame=single,
      basicstyle=\scriptsize\ttfamily,
      commentstyle=\scriptsize\itshape\color{gray!60!black},
      morekeywords={await,while},
      keywordstyle=\bfseries\color{blue!60!black},
      columns=fullflexible,breaklines=false,
      xleftmargin=0pt,xrightmargin=0pt,
      framexleftmargin=0pt,framexrightmargin=0pt,
      numbers=left,numbersep=5pt,
      numberstyle=\tiny\color{gray}]
prepare_shared(s, w)
while (pre_s == 0) {
  prime(reference)
  access(s)
  access(w)
}
# synchronize
pre_s = 0
await s == 1
r = reference_op() # marker 1
w = r
    \end{lstlisting}
  \end{minipage}
  \caption{High-level pseudocode for unpadded \AttackName. \code{prepare\_shared} normalizes the cache and coherence state of $s$ and $w$; \code{target\_op} and \code{reference\_op} propagate distinct marker values through data-dependent result stores. We omit the reset logic between measurements for clarity.}
  \label{fig:pseudocode-unpadded}
\end{figure}

\Cref{fig:pseudocode-unpadded} shows the implementation at a high level.
The reference thread continuously prepares the reference operand and the shared state while waiting for a request from \tprobe.
A short pre-synchronization handshake then moves \tref into a tight spin loop on $s$, reducing variation in its response path.
After \tprobe writes to $s$, both threads execute their respective operations and propagate distinct marker values to $w$.
A separate completion flag (omitted here) ensures that \tprobe reads $w$ only after \tref has issued its result store.
For native experiments, we implement both threads entirely in assembly and expose bindings to higher-level languages.
We provide implementations for x86\_64, AArch64, RISC-V64, and LoongArch64.
The implementations differ primarily in assembly syntax and memory-ordering details.
The primitive itself requires only loads and stores, concurrent execution, and shared memory.
For padded \AttackName, we generate $\Delta_{pad}$ with a chain of dependent register-only additions whose final value feeds the result store.
Varying the chain length provides fine-grained control over the padding without relying on a timer.

\subsection{Threat Model}\label{subsec:threat-model}

We assume an attacker with unprivileged code execution on the target machine.
\AttackName inherently requires (i) two threads that can execute concurrently, (ii) writable shared memory for synchronization, completion, and the result word, and (iii) the ability to invoke the target and reference operations.
It does not require a timestamp source, performance counter, privileged instruction, or ISA-specific extension.
Our evaluated native implementation also assumes the attacker can keep the two threads on stable logical cores, e.g., through the affinity interface available to unprivileged users on Linux and Windows.
Without explicit affinity control, scheduler migration increases $\Delta_{sync}$ variability and may require repeated measurements, but this does not change the construction.
Our padded implementation generates dependency chains at runtime for convenience.
However, runtime code generation is not necessary: pre-generated chains or a dependency-preserving loop can provide the same controlled delay.

\section{Evaluation}\label{sec:evaluation}

We evaluate four properties of \AttackName: the synchronization precision that bounds its temporal resolution (\Cref{subsec:delta-sync}), its single-shot accuracy when classifying cache states (\Cref{subsec:measuring-memory-access-latencies}), its performance relative to counting-thread timers (\Cref{subsec:comparison-to-counter-threads}), and its applicability inside a web browser (\Cref{subsec:attackname-in-the-browser}).
Our test set contains \PlatformCount physical CPUs spanning x86\_64, AArch64, RISC-V64, and LoongArch64.
Seven of these CPUs contain heterogeneous core types.
Treating each core type as a separate platform yields \num{45} native platform configurations.
\Cref{tab:precision,tab:precision_ht,tab:precision_wasm} in \Cref{sec:precision-analysis} report full per-platform results.
Unless stated otherwise, we pin \tprobe to logical core 0 and \tref to the next physical core.
On the \num{20} configurations supporting simultaneous multithreading (SMT), we repeat each experiment with the threads on sibling logical cores.
Where possible, we disable hardware prefetchers.
We express short delays in \emph{\texttt{add}-cycles}, defined as the latency of one register-only \texttt{add} in a dependency chain.
This unit is portable across our four ISAs and avoids conflating core cycles with fixed-frequency timestamp-counter ticks.
On many microarchitectures, one \texttt{add}-cycle corresponds to one core cycle~\cite{abel2019uops,Intel_combined}.

\subsection{Synchronization Precision}\label{subsec:delta-sync}

The resolution of unpadded \AttackName depends on the synchronization latency \rdelta{sync} and, more importantly for padded \AttackName, on its run-to-run jitter.
Thus, we measure both with respect to thread placement.

\begin{figure}[t]
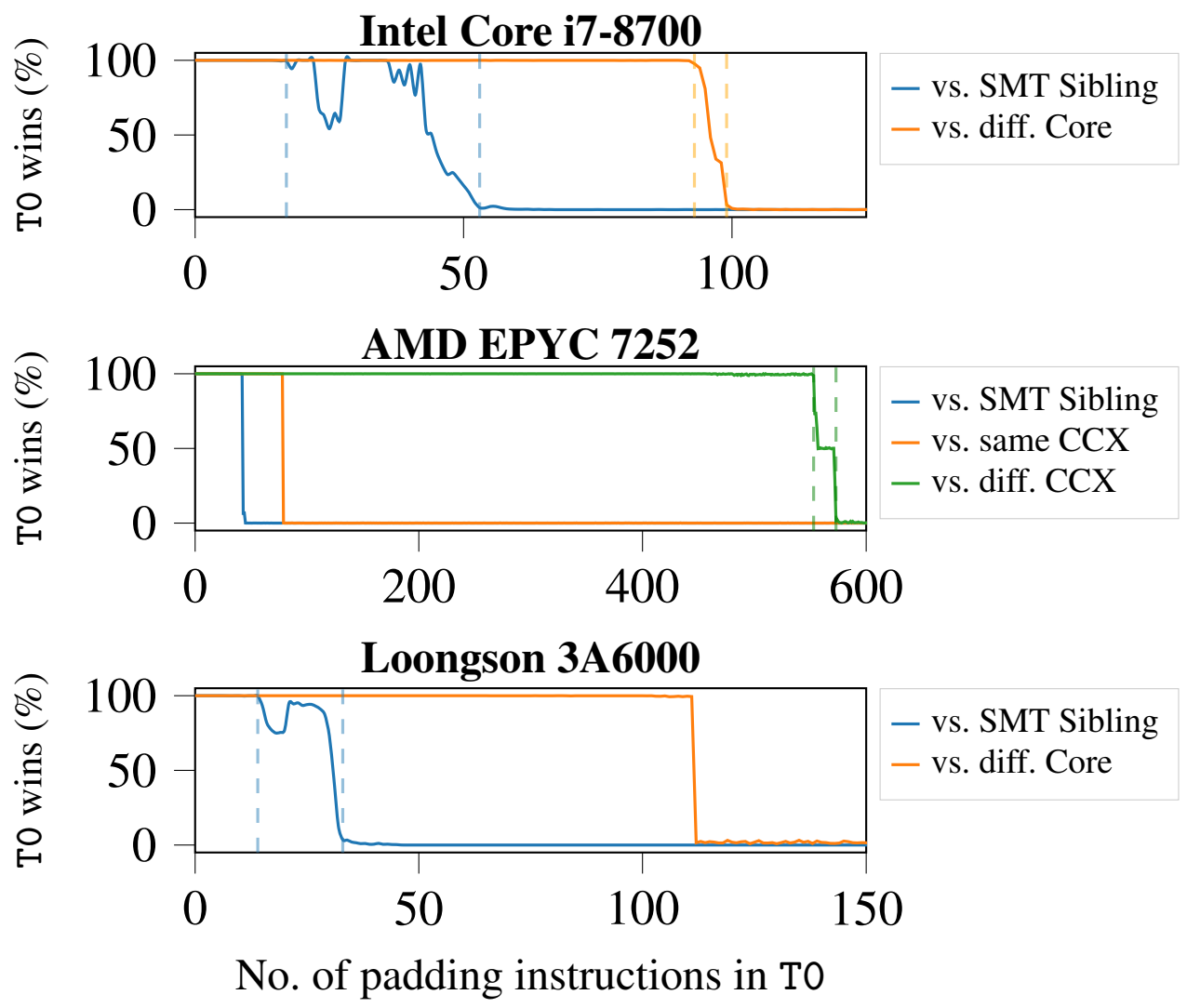

  \flushleft
  \newcommand{\synctimeheight}[0]{2.75cm}
\newcommand{\synctimewidth}[0]{.75\linewidth}
  \input{plots/sync_time_Intel-Core-i7-8700.tikz}
  \input{plots/sync_time_AMD-EPYC-7252.tikz}
  \input{plots/sync_time_Loongson-3a600.tikz}
  \caption{Synchronization transition on representative CPUs ($n=2048$ races per padding value). Increasing the dependent padding in \tprobe moves the winner from \tprobe to \tref. The transition midpoint estimates \rdelta{sync}. The shaded uncertain range captures values for which neither thread wins at least \SI{95}{\percent} of races. Core placement affects both quantities.}
  \label{fig:sync-time}
\end{figure}

\paragrabf{Method}
To measure \rdelta{sync}, we instantiate a padded race without target or reference memory accesses.
After \tprobe releases \tref through the synchronization word, \tprobe first executes a dependency chain of $d$ register-only \texttt{add} instructions, while \tref writes after observing the release.
Thus, the result compares the controlled padding $d$ against \rdelta{sync}.
We execute \num{2048} such races for each $d \in [0, 1000)$ and record the probability $p(d)$ that \tprobe wins.
At $d=0$, \tprobe normally wins, and as $d$ increases, the winner transitions to \tref.
On most CPUs, this transition zone consists of multiple $d$ values where both threads have a high chance of winning.
This is illustrated in \Cref{fig:sync-time}.
We call this the \emph{uncertain range} $U$, and define the lower endpoint as the largest $d$ for which $p(d)\geq\SI{95}{\percent}$, and its upper endpoint as the smallest $d$ for which $p(d)\leq\SI{5}{\percent}$, i.e., $U$ brackets the transition.
Its midpoint estimates \rdelta{sync}, while its width captures the observable synchronization jitter.
The upper edge of $U$ provides a conservative bound for unpadded \AttackName, whose condition is $\Delta_{slow}-\Delta_{fast}>\Delta_{sync}$.
For padded \AttackName, the width $|U|$ approximates the minimum latency separation that can be classified reliably with one race.
Smaller differences can still be recovered by combining repeated measurements, as we demonstrate in \Cref{subsec:determining-llc-slices}.
We run this experiment on all \num{45} platform configurations.
To study topology effects, we keep \tprobe on core 0 and place \tref on every other logical core.
All reported values of \rdelta{sync} are midpoints of the corresponding uncertain ranges.

\paragrabf{Cross-Core versus SMT Placement}
SMT siblings provide a shorter synchronization path because both threads share the private cache hierarchy.
However, they also share frontend and execution resources, and we observe a substantially wider uncertain range on most SMT configurations.
Thus, SMT reduces the mean synchronization latency but generally worsens single-shot temporal precision.
For the remainder of the paper, we prefer separate physical cores unless stated otherwise.
Across separate physical cores, the exact core pair has little effect on many CPUs.
The measured \rdelta{sync} is identical across placements on \num{8} platforms and varies by fewer than \num{10} \texttt{add}-cycles on another \num{21}.
The main exception is topology that crosses an internal interconnect boundary.
For example, on the AMD \labthirtyCPU, \rdelta{sync} is \num{78} \texttt{add}-cycles when both threads reside in the same core complex (CCX), but averages \num{563} \texttt{add}-cycles across CCXs.
We observe the same behavior on all tested multi-CCX AMD systems: selecting cores within one CCX significantly improves the race boundary.
On several platforms, $U$ is only one \texttt{add}-cycle wide, the finest granularity our setup resolves.
Repeated experiments on these systems distinguish chains differing by one dependent \texttt{add}.
While this result characterizes the synchronization path, variability in the target operation can widen the effective transition during an attack.
Accordingly, the values in \Cref{sec:precision-analysis} are optimistic lower bounds on the total classification noise.

\paragrabf{Predicting L1-versus-DRAM Feasibility}
Knowledge of \rdelta{sync} enables predicting the success of, e.g., a cache-state classifier.
To demonstrate this, we sweep \rdelta{pad} from \num{0} to \num{1000} \texttt{add}-cycles for both an L1 hit and a DRAM access, taking \num{2048} samples per case and padding value.
We conservatively estimate the difference between an L1 hit and a DRAM access $\Delta^{L1}_{DRAM}$ as the separation between the two transition regions: the highest padding for which the L1 case still wins more than \SI{5}{\percent} of races and the lowest padding for which the DRAM case wins fewer than \SI{95}{\percent}.
When running on core 0 and a separate physical core, $\Delta^{L1}_{DRAM}$ exceeds \rdelta{sync} on all but two platforms, \labfourtyfourCPU and \labeightythreeCPU.
The unpadded condition therefore predicts successful L1-versus-DRAM classification on the remaining platforms.
In contrast, $\Delta^{L1}_{DRAM}$ exceeds $|U|$ on every platform, predicting that calibrated padding can separate the two states throughout the test set.
The direct classification experiments below, together with the AES case study in \Cref{subsec:aes-t-table}, confirm the padded prediction on every platform.
Both platforms predicted to fail without padding indeed do, although two further platforms fall short for reasons we discuss below.

\subsection{Cache-Latency Classification}\label{subsec:measuring-memory-access-latencies}

In this section, we evaluate \AttackName as a primitive for inferring a cache line's state.
We consider L1 hits versus DRAM accesses, common in \FlushReload-style attacks~\cite{Yarom2014Flush}, and LLC hits versus DRAM accesses, as required by LLC \PrimeProbe~\cite{Liu2015Last}.
For padded \AttackName, we additionally test the smaller L1-versus-L3 and L1-versus-L2 differences.

\paragrabf{Method}
For every combination of cache state for the address and padding value, we collect $2^{14}$ samples from each class, with \rdelta{pad} from \num{0} to \num{500} \texttt{add}-cycles.
Note that padding zero is the unpadded construction.
We report the $F_1$ score with the fast state as the positive class.
We prepare cache levels using physical-address-based eviction sets, as detailed in \Cref{sec:reliable-cache-eviction}.
Where a sufficiently precise timer is available, we independently verify that eviction leaves the target at the intended cache level.
We cannot perform this validation on the RISC-V and LoongArch64 systems or on \labseventyoneCPU.
Thus, their L2/L3 results are conservative: a low score may indicate imperfect state preparation rather than insufficient resolution of \AttackName.
For LLC-versus-DRAM measurements on SMT siblings, \tref also accesses an LLC-resident reference line.
With separate physical cores, we instead retain an L1-resident reference because \rdelta{sync} already shifts the reference path beyond typical LLC latency.
Avoiding an LLC access in \tref removes interconnect-induced jitter~\cite{paccagnella2021lord} and improves the decision boundary.

\paragrabf{Unpadded \AttackName}
For L1 versus DRAM on separate physical cores, \num{41} of \num{45} platforms achieve $F_1\geq0.97$.
The four exceptions expose two distinct failure modes.
On \labseventyoneCPU and the efficiency cores of \labninetyfiveCPU, DRAM latency varies enough that some nominally slow probes complete before the reference path.
On \labfourtyfourCPU and \labeightythreeCPU, the L1-to-DRAM separation is too small relative to \rdelta{sync}, violating the unpadded condition.
Among all configurations, \num{38} produce fewer than \num{20} errors across \num{32768} classified accesses, and \num{4} classify every access correctly.
All tested RISC-V and LoongArch64 platforms achieve high scores, showing that the unpadded primitive supports timer-free, calibration-free \FlushReload-style attacks beyond x86 and ARM.
On SMT siblings, all \num{20} platforms exceed $F_1=0.99$ for L1 versus DRAM, with \num{18} platforms having fewer than \num{20} errors and \num{8} platforms being perfect.
The shorter synchronization path dominates the additional SMT jitter for this large latency difference.
For LLC versus DRAM, the cross-core configuration again performs strongly: \num{37} of \num{45} platforms exceed $F_1=0.99$, including \num{5} perfect classifiers.
These platforms support an unpadded, calibration-free LLC \PrimeProbe readout.
On SMT siblings, \num{6} of \num{20} platforms exceed $F_1=0.99$.
Here, the smaller cache-latency separation is more sensitive to the wider uncertain range, confirming the placement result from \Cref{subsec:delta-sync}.

\begin{figure*}[t]
  \centering
  \input{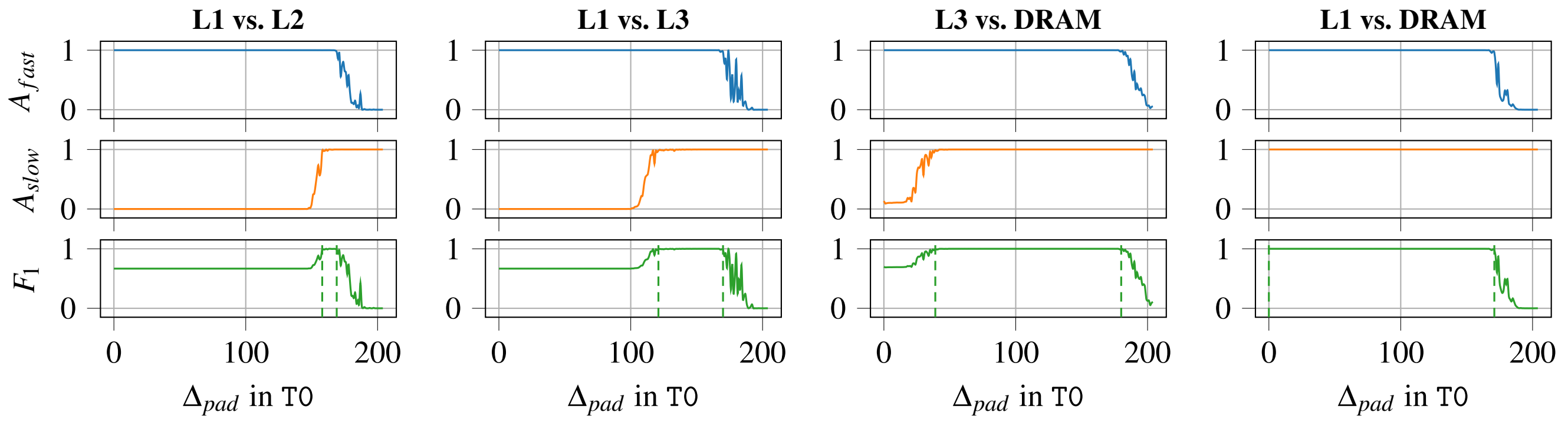}
  \caption{Cache-state classification on \labthirtythreeCPU with cross-core placement ($n=16384$ per class and padding value). The fast-state accuracy $A_{fast}$ decreases at a larger padding value than the slow-state accuracy $A_{slow}$. Their overlap forms the golden range $G$, in which the combined classifier achieves $F_1>0.98$.}
  \label{fig:padding-results}
\end{figure*}

\paragrabf{Padded \AttackName}
Padding converts a single race boundary into a tunable classifier.
\Cref{fig:padding-results} illustrates how padding allows tuning the classifier on a sample \labthirtythreeCPU.
At zero padding, both the fast and slow target may finish before \tref, resulting in an unusable classifier with $F_1\approx 2/3$.
As padding increases, the slow state crosses the reference boundary first and the fast state later.
Between these transitions, both classes are classified accurately.
We define the \emph{golden range} $G$ as a contiguous interval of padding values for which $F_1>0.98$.
Its width measures calibration tolerance: a wider $G$ makes the classifier less sensitive to small shifts in execution conditions.
Every platform has a non-empty golden range for L1 versus DRAM\@.
Thus, padded \AttackName provides timer-free \FlushReload-style classification on all tested CPUs, including platforms where the unpadded condition fails.
For LLC versus DRAM, \num{42} of \num{45} platforms have a golden range.
The finer L1-versus-L3 difference is separable with one race on \num{32} platforms, and even L1 versus L2 is separable on \num{15}.
The latter includes the efficiency cores of \labseventytwoCPU and the Apple M1.
On the M1, the achieved granularity significantly exceeds that of the unprivileged architectural timer~\cite{ARM2021GenericTimer,Lipp2016}.
With SMT placement, all \num{20} platforms retain a golden range for L1 versus DRAM.
Smaller differences are less robust: \num{13} platforms expose a golden range for L1 versus L3, and \num{7} for LLC versus DRAM.
This confirms the tradeoff established in \Cref{subsec:delta-sync}, i.e., SMT minimizes \rdelta{sync}, whereas separate physical cores usually minimize the transition width.

\takeaway{Unpadded \AttackName provides calibration-free L1-versus-DRAM classification on \num{41} of \num{45} cross-core platforms. Calibrated padding extends coverage to all tested CPUs and can even classify adjacent cache levels.}

\subsection{Comparison with Counting Threads}\label{subsec:comparison-to-counter-threads}

Counting threads are the standard timing fallback when an explicit high-resolution timer is unavailable~\cite{Wray1992,Lipp2016,Schwarz2017Timers,Gras2017aslr,Schwarz2020malware,Hetterich2022Branch}.
They require the same basic resources as \AttackName, i.e., parallel execution and shared memory, but expose time through periodically visible counter updates.
A cache attack must then calibrate a numerical threshold over this emulated clock.
We compare the two approaches first at the level of temporal resolution and then through direct cache-state classification.

\paragrabf{Counting-Thread Implementation}
Our assembly implementation is a loop that increments a register and stores it to shared memory, requiring three instructions per iteration.
We evaluate all \num{45} native platform configurations cross-core and the \num{20} SMT configurations on sibling threads.

\paragrabf{Effective Counter Resolution}
The relevant resolution is not the local loop throughput but the interval between counter values observable by another core.
Store buffering, cache ownership, and coherence propagation can combine or delay updates.
We call the mean interval between globally observed changes the \emph{edge width}.
A branch-free observer repeatedly loads the counter and counts value changes.
We measure the wall-clock time required to observe \num{10000000} changes using a coarse OS timer, then convert it to \texttt{add}-cycles by benchmarking \num{65536} repetitions of a \num{1024}-instruction dependent-add chain.
The observer introduces a small positive bias, but the counter loop itself takes fewer than \num{7} \texttt{add}-cycles per iteration on every platform, whereas the measured edge width is substantially larger on nearly all of them.
The cross-core edge width is higher than \rdelta{sync} by more than \num{30} \texttt{add}-cycles on \num{28} of \num{45} platforms, while the opposite occurs on only \num{5}.
Thus, the unpadded race usually has a finer boundary than a cross-core counting thread.
On SMT siblings, edge width and \rdelta{sync} are typically within \num{15} \texttt{add}-cycles, although \labninetyoneCPU and \labseventynineCPU are outliers at \num{3} and \num{249} \texttt{add}-cycles, respectively.
For padded \AttackName, the relevant quantity is $|U|$ rather than \rdelta{sync}.
On \num{19} of \num{20} platforms, the cross-core uncertain range is narrower than the edge width of an SMT-sibling counting thread, indicating a substantially finer calibrated decision boundary.

\paragrabf{Direct Classification}
We use the counting thread as a timer for the same L1-versus-DRAM and LLC-versus-DRAM classifications.
For each pair, we profile the mean fast and slow readings, set the threshold to
$T_{fast}+\frac{1}{2}(T_{slow}-T_{fast})$, and classify the fast and slow cases over \num{65536} iterations.
In the cross-core setting, counting threads exceed $F_1=0.97$ on \num{36} of \num{40} platforms for L1 versus DRAM and \num{32} for LLC versus DRAM.
On the same subset, unpadded \AttackName reaches \num{37} and \num{33}, respectively, despite requiring no numerical threshold.
Padded \AttackName reaches all \num{40} platforms for L1 versus DRAM and \num{35} for LLC versus DRAM.
Thus, cross-core \AttackName matches or slightly exceeds the counting-thread baseline, and padding further increases coverage.
The SMT result is more nuanced.
Counting threads exceed $F_1=0.97$ on \num{16} of \num{18} platforms for both latency pairs.
Unpadded \AttackName reaches all \num{18} for L1 versus DRAM but only \num{7} for LLC versus DRAM. Padded \AttackName exceeds $F_1=0.98$ on all \num{18} and \num{9}, respectively.
Consequently, an SMT-sibling counting thread remains the stronger baseline for the smaller LLC-to-DRAM difference, but at the cost of threshold calibration.

\takeaway{Across physical cores, \AttackName is at least as accurate as counting-thread timing while unpadded \AttackName eliminates threshold calibration. Padded \AttackName provides the broadest coverage.}

\subsection{Browser Execution}\label{subsec:attackname-in-the-browser}

We demonstrate that \AttackName in WebAssembly is a practical alternative to counting threads in browsers.
Browsers are an important restricted environment for microarchitectural attacks because they execute untrusted code, support parallel execution through Web Workers, and expose shared memory through \texttt{SharedArrayBuffer} under cross-origin isolation~\cite{Oren2015,Roettger2021SpectreJS,W3C2015WebWorker,FirefoxSharedArrayBuffer}.

\paragrabf{Implementation and Scope}
We implement both threads in WebAssembly and execute them in Chromium v147.0 on x86\_64.
Using JavaScript, we rewrite the WebAssembly bytecode to insert a variable-length chain of dependent additions for \rdelta{pad}.
Inspection of the generated machine code confirms that Chromium emits the intended register-only dependency chain.
The primitive itself uses only Web Workers and \texttt{SharedArrayBuffer}.
Standard browser code cannot execute \texttt{clflush}, and this experiment requires known cache states to measure classification accuracy.
Thus, we use a patched Chromium build to prepare and label the probe line through native cache-management instructions.
We do not use these hooks for synchronization, padding, or readout, and the WebAssembly primitive does not normalize the coherence state of $s$ and $w$ as the native implementation does.
We repeat the synchronization and L1-versus-DRAM experiments with the same cross-core and SMT placements as in native code, again treating heterogeneous core types separately.
Per padding value, we collect \num{2048} synchronization races and \num{4000} cache-state samples.

\paragrabf{Results}
The WebAssembly and native results are strongly correlated (Pearson $r=0.930$ across the matched configurations).
Browser execution increases the mean \rdelta{sync} from \num{98.7} to \num{129} \texttt{add}-cycles and widens the mean uncertain range from \num{23.7} to \num{37.8} \texttt{add}-cycles.
The additional latency and jitter are consistent with the longer generated response path and the missing coherence-state preparation for $s$ and $w$.
Despite this reduced resolution, unpadded WebAssembly \AttackName obtains $F_1\geq0.99$ in \num{46} of \num{50} configurations for L1 versus DRAM.
Padded \AttackName yields a golden range on all \num{50} configurations.
Its mean width is \num{84.8} \texttt{add}-cycles, close to the native mean of \num{86.9}, and the per-configuration widths remain strongly correlated (Pearson $r=0.853$).
Thus, the shared-memory race survives browser compilation with only a moderate increase in synchronization noise.

\takeaway{WebAssembly preserves the behavior and accuracy of \AttackName: padded classification succeeds in every tested browser configuration.}

\section{Case Studies}\label{sec:case-studies}
In this section, we show that \AttackName can substitute for a timer in \num{5} case studies.
\Cref{subsec:aes-t-table} demonstrates \AttackName in a timerless attack against AES\@.
In \Cref{subsec:timerless-evict-reload}, we show that \AttackName can find minimal eviction sets, and use them for an Evict+Reload-based covert channel on an Apple M1 CPU\@.
\Cref{subsec:determining-llc-slices} demonstrates that \AttackName can detect the LLC slice of an address with aggregated measurements.
\Cref{subsec:other-latency-types} shows that \AttackName also resolves non-cache latencies.
Finally, in \Cref{subsec:attackname-in-lua}, we show that \AttackName does not strictly require access to binary code by implementing the primitive in Lua.

\subsection{AES T-Table Attack}\label{subsec:aes-t-table}

\pgfplotsset{
    colormap={whiteblack}{
        rgb(0cm)=(1,1,1);
        rgb(1cm)=(0,0,0);
    }
}

\newcommand{\ttablematrix}[2]{
        \nextgroupplot[
            title={#2},
            title style={at={(axis description cs:.5,.91)}, anchor=south, font={\bfseries\selectfont\footnotesize}},
            height=4cm,
            width=4cm,
            xlabel=Plaintext Byte,
            xtick={0, 4, 8, 12, 15},
            xticklabels ={00, 40, 80, c0, f0},
            xticklabel style={font={\fontfamily{cmtt}\selectfont\footnotesize}},
            xlabel style={at={(axis description cs:.5,-.3)}, anchor=south, font={\scriptsize}},
            ytick={0, 4, 8, 12, 15},
            yticklabels ={0, 4, 8, 12, 15},
            yticklabel style={font={\fontfamily{cmtt}\selectfont\footnotesize}},
            ylabel style={at={(axis description cs:-.175,0.5)}, anchor=south, font={\scriptsize}},
            ylabel=Cache Line,
            xmin=-.5,xmax=15.5,
            ymin=-.5,ymax=15.5,
            enlargelimits=false,
            colormap name=whiteblack,
            axis on top,
        ]
            \addplot[
            matrix plot*,
            point meta=explicit,
            ] file {#1};
}

\begin{figure}[t]
    \centering
    \begin{tikzpicture}
        \begin{groupplot}[
            group style={group size=2 by 2, horizontal sep=1.25cm, vertical sep=1.25cm,},
        ]
        \ttablematrix{data/t-table/lab89.txt}{Intel \labeightynineCPU}
        \ttablematrix{data/t-table/rpi3b.txt}{\labninetyeightCPU}
        \ttablematrix{data/t-table/lab53.txt}{\labfiftythreeCPU}
        \ttablematrix{data/t-table/lab91.txt}{\labninetyoneCPU}
        \end{groupplot}
    \end{tikzpicture}
    \caption{A selection of heatmaps produced by the T-table attack using unpadded \AttackName on \num{4} different CPUs. All show the diagonal line expected when the first key byte is 0.}
    \label{fig:t-table-matrices}
\end{figure}
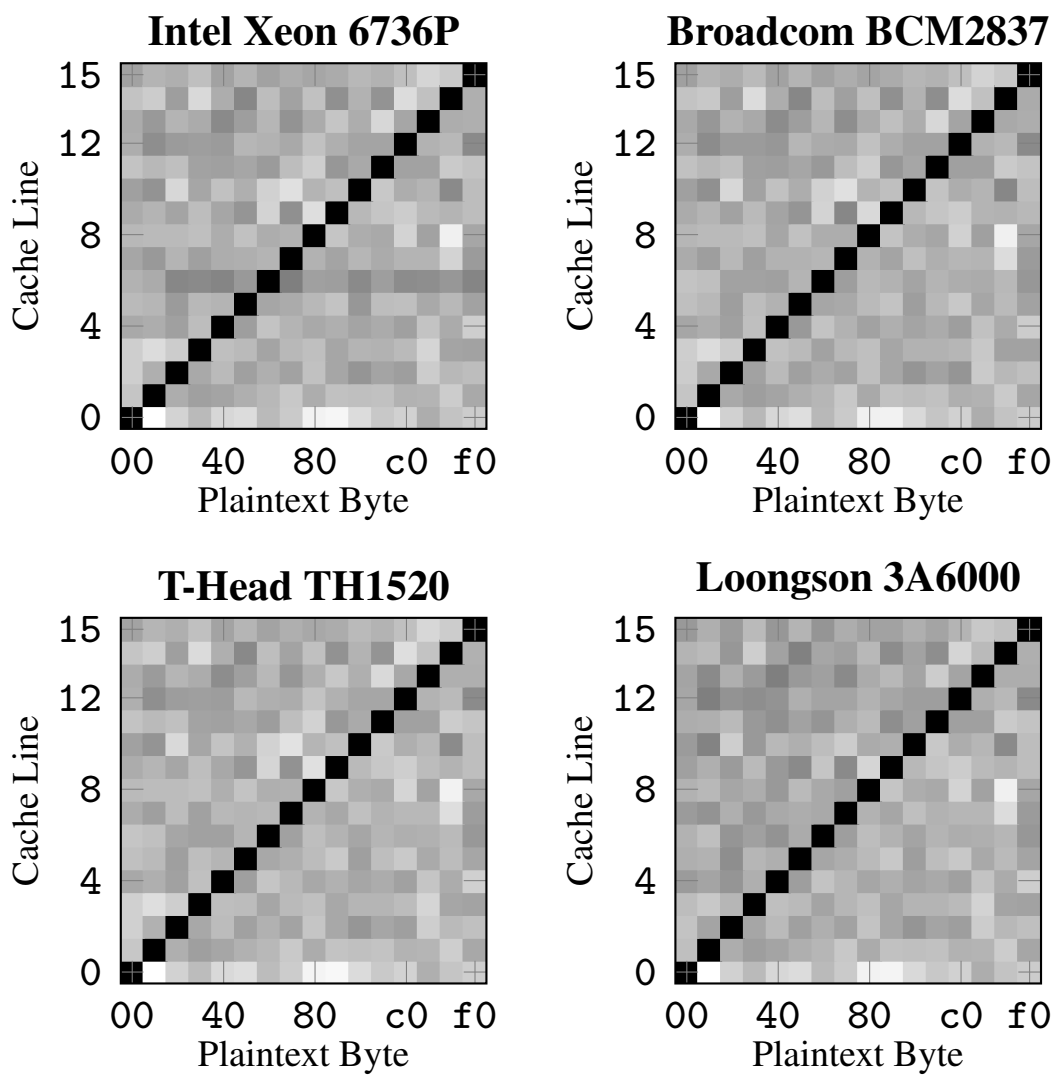

To evaluate \AttackName's performance in a real-world attack scenario, we use it in a Flush+Reload-style attack targeting OpenSSL's T-table-based AES implementation.
Specifically, we mount a first-round chosen plaintext attack to partially recover the first key byte from memory access patterns~\cite{Osvik2006}.
This attack has emerged as a standard benchmark for cache attacks, making our results directly comparable to a large body of existing work~\cite{Gruss2016Flush,Disselkoen2017prime,VanSchaik2018malicious,Briongos2020reload,Lipp2020takeaway,Purnal2021PrimeScope,Gerlach2023A,Purnal2023showtime,Gerlach2024Hash,Weber2024no,Thomas2025Exfilstate}.

\paragrabf{Setup}
We specifically target OpenSSL v.3.5.0, compiled for all \num{4} architectures in our evaluation set using GCC v15.3.
OpenSSL's T-table AES implementation uses no architecture-specific code, enabling us to mount the experiment on all \PlatformCount CPUs in our evaluation set.
However, we exclude Apple's M1 and M2 architectures, as they feature a cache line size of \SI{128}{\byte} at the point of coherence, whereas the attack requires \SI{64}{\byte}.
We run unpadded \AttackName on separate physical cores, with \tprobe on core 0 and \tref on the physical core with the next-highest ID.
As with the experiments in \Cref{sec:evaluation}, we treat CPUs with hybrid architectures as two test platforms.
With \num{5} of our \num{36} remaining CPUs being hybrids, this results in \num{41} total platforms.
Since the memory containing the T-tables is physically shared between the attacker and victim, we can use \AttackName in a Flush+Reload-like attack.
We flush each relevant cache line to DRAM, perform the encryption, and then probe the cache lines using \AttackName.
This corresponds to the L1 vs.\ DRAM configuration tested in \Cref{subsec:measuring-memory-access-latencies}.
On platforms without a flush instruction, we evict cache lines using the method described in \Cref{sec:reliable-cache-eviction}.
In each iteration of the attack, we twice encrypt a chosen 128-bit plaintext block with \num{0} as the first key byte.
We set the first byte of this block to a multiple of 16, with the rest containing random data generated from a fixed seed.
Overall, we run \num{1000} probing iterations for each possible setting of the first byte, probing all \num{16} relevant cache lines in each iteration.
We analyze the resulting access patterns to determine whether the attack is successful and repeat the attack \num{1000} times per platform to gauge its reliability.

\paragrabf{Results}
Of the \num{41} tested platforms, \num{37} have a success rate of \SI{100}{\percent}.
One platform, the \labnineCPU, reaches \SI{99.9}{\percent}.
\Cref{fig:t-table-matrices} shows a representative sample of the heatmaps the attack yields.
All clearly show the diagonal line expected when the first key byte is 0.
Moreover, the noise patterns in non-diagonal cells show similarities across machines, indicating high probing accuracy.
On average across all platforms, the median runtime is \SI{315}{\milli\second}.
The lowest average runtime is \SI{218}{\milli\second} on the AMD \labeightyCPU, and the highest is \SI{17.4}{\second} on the \labfourtyeightCPU.
There are only \num{3} machines on which the attack is unreliable: the \labthirtytwoCPU (\SI{3.5}{\percent} success rate), the \labeightythreeCPU (\SI{55}{\percent} success rate), and the Intel \labfourtyfourCPU (\SI{0}{\percent} success rate).
For the latter two, this is expected, as \Cref{subsec:measuring-memory-access-latencies} predicts: on these platforms $\Delta^{L1}_{DRAM}$ does not exceed \rdelta{sync}, so the precondition for the unpadded variant does not hold.
We confirm that a padding of \num{75} \texttt{add}-cycles restores a \SI{100}{\percent} success rate.
For the \labthirtytwoCPU, which has only two parallel threads, we believe the low success rate results from its small cache (\SI{2}{\mega\byte} LLC) combined with its high susceptibility to outside interference due to its low core count.

\subsection{Evict+Reload}\label{subsec:timerless-evict-reload}

In addition to demonstrating \AttackName in a Flush+Reload attack, we show that attackers can use it to find eviction sets without a precise timer.
For this, we implement a cross-core covert channel on an Apple M1 system using Evict+Reload on the L2 cache.
On this architecture, traditional timing-based cache attacks are challenging since only a coarse-grained \SI{24}{\mega{}Hz} timer (CNTVCT) is available.

\paragrabf{Finding Minimal Eviction Sets}
The M1 processor uses a complex hash function to index its L2 cache~\cite{Yu2023Synchronization}.
Its hash depends on physical address bits outside the page offset, so we cannot compute minimal eviction sets from virtual addresses alone.
Instead, we start with an excessive eviction set of \num{1536} addresses that have the same page offset as the victim address, \ie a \SI{24}{\mega\byte} array.
We then remove one address at a time and use the padded \AttackName primitive to detect whether the eviction set can still evict the target address from the L2 cache.
If eviction still succeeds, we remove this address from the eviction set, and repeat until we reach \num{12} addresses, i.e., the L2 cache's associativity.
While we only use minimal eviction sets in our experiment, note that minimality is not a strict requirement.
The initial set of \num{1536} could also be used with a potential reduction in throughput.

\paragrabf{Synchronization}
We use the coarse-grained \texttt{CNTVCT} timer to lay out fixed time slots for each transmitted bit.
To signal the start of a transmission, the sender holds the one-signal for eight consecutive time slots, which the receiver decodes as a start byte of \code{0xff}. The receiver discards all bytes until it observes this value.
If no coarse-grained timer is available, an additional cache line could transmit a clock signal, or post-processing could align the measurement to bits.

\paragrabf{Receiver}
The receiver uses two threads with the \AttackName primitive in a loop.
Since \rdelta{sync} is sufficiently long, we use an L1 cache hit and additional padding instructions instead of an L2 cache hit in the reference thread.
After each measurement, we evict the shared cache line from the L2 cache using a minimal eviction set.
A counter is initialized to 0 at the start of a time slot.
On each cache hit, it is incremented; on a cache miss, it is decremented.
If the counter is positive at the end of a time slot, the received value is 1; otherwise, it is 0.

\paragrabf{Sender}
To send a 1, the sender repeatedly accesses the shared cache line in a loop.
To send a 0, the sender evicts the shared cache line from its L1 cache.
This is necessary since eviction by the receiver from the L2 cache can fail if the value is still present in the L1 cache of a different core.

\paragrabf{Results}
We run the covert channel on the M1's Firestorm performance cores at \SI{30}{\kilo\byte/\second} and transmit \SI{64}{\kilo\byte} of data, conducting \num{200} runs in total.
We find a minimal eviction set in \num{163} of these runs (\SI{81.50}{\percent}) with a median construction runtime of \SI{289}{\milli\second}.
If an eviction set is found, a median of \SI{96.87}{\percent} of bits is received correctly without error correction.
This shows that the \AttackName primitive can reliably conduct eviction-based cache side-channel attacks on platforms without highly accurate timers.

\subsection{Determining LLC Slices}\label{subsec:determining-llc-slices}

\begin{figure}[t]
  \input{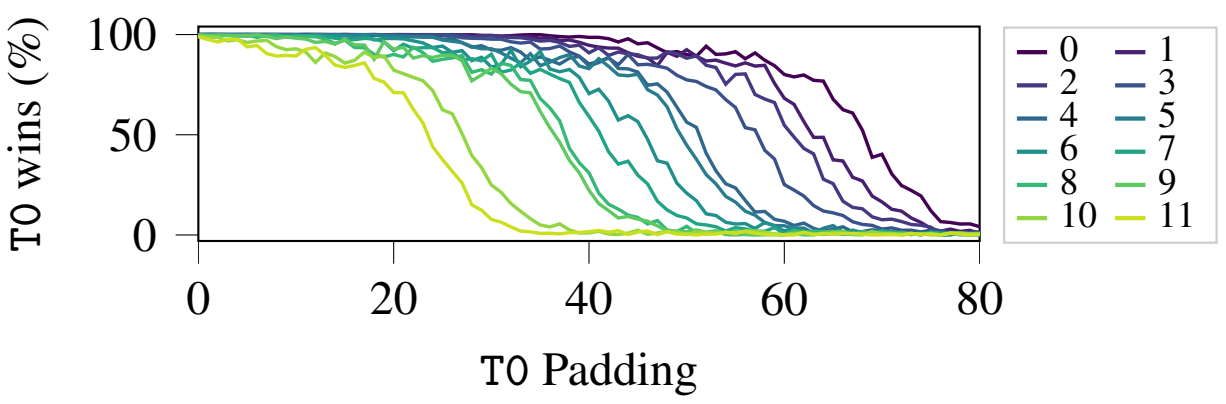}
  \caption{Win rate of \tprobe with the probe address in different LLC slices of an Intel \labfourtytwoCPU ($n = 4096$ per sample). The threads run on cores 0 and 1, which are SMT siblings.}
  \label{fig:llc-slices}
\end{figure}
The LLCs of multicore Intel CPUs are segmented into \textit{slices}, each typically grouped with one physical CPU core~\cite{Irazoqui2015slice,Gerlach2024Hash,Rainer2025slice,Morgan2025slice}.
Slices are interconnected in a ring- or mesh-like routing structure, and LLC access latency depends on how far a slice is from the core issuing the operation.
A complex microarchitectural hash function over the physical address bits determines which slice an address uses.
We show that with multiple aggregated measurements, \AttackName can determine an address's LLC slice.
For this experiment, we use the Intel \labfourtytwoCPU, which features \num{12} LLC slices on a bidirectional ring interconnect.
The hash function on this CPU is known~\cite{Rainer2025slice}, giving us reliable ground truth.
To reliably evict addresses to the LLC, we use the eviction strategy outlined in \Cref{sec:reliable-cache-eviction}.
We first measure \tprobe's win rate at different padding levels, with the probe address in different LLC slices and the threads on SMT siblings.
The results of this experiment are shown in \Cref{fig:llc-slices}.
We observe that the point at which the win rate drops below \SI{50}{\percent} is different for each LLC slice, with the relative differences varying from \num{45} \texttt{add}-cycles for slices \num{0} and \num{11}  to just \num{1} \texttt{add}-cycle for slices \num{8} and \num{9}.
However, we also observe that no padding level creates a scenario where, e.g., an address from slice \num{8} always causes \tprobe to win and an address from slice \num{9} always causes it to lose.
To determine the cache slice, we therefore use the location of the win rate's \SI{50}{\percent} threshold as a stable marker.
Hence, we must take multiple measurements per test address with different padding levels.
We measure \AttackName's performance for determining an address's cache slice in a separate experiment.
It consists of two steps: A \emph{calibration phase}, which determines the \SI{50}{\percent} thresholds of the previous experiment using addresses with known cache slices, and an \emph{evaluation phase}, which determines the cache slice of \num{1000} random addresses and compares it to the ground truth.
In the calibration phase, we determine the \SI{50}{\percent} thresholds with the threads pinned to SMT sibling cores \num{0} and \num{1} and cores \num{14} and \num{15}, giving us measurements from two points on the ring.
The evaluation phase mirrors this, determining the \SI{50}{\percent} threshold of a test address in both configurations, using \num{512} measurements per padding level.
To determine the slice, we treat the calibration thresholds and the two test-address thresholds as points in a two-dimensional space, and choose the slice whose calibration point is closest to the observed test-address thresholds.
Using this technique, we observe a slice detection accuracy of \SI{96.9}{\percent} with \num{1000} random addresses.
This means that \AttackName can distinguish differences on the order of individual cycles in scenarios where multiple measurements are permitted.
The uncertain range for SMT siblings of this CPU in \Cref{subsec:delta-sync} is \num{31} \texttt{add}-cycles wide, confirming that multiple measurements can further improve the temporal resolution.

\subsection{Beyond Cache-Access Latencies}\label{subsec:other-latency-types}

While this paper focuses mostly on cache access latencies, \AttackName can, in principle, distinguish any two microarchitectural timing differences.
We also show it can profile other stateful components, such as the TLB and parts of the execution engine.

\paragrabf{AVX Wakeup Throttling}
On many Intel CPUs, 256-bit AVX instructions executed after a prolonged period of non-AVX usage are subject to a transient performance penalty.
In this ``cold'' state, latencies of AVX instructions are significantly higher~\cite{Schwarz2019netspectre}.
Haj-Yahya \etal show that this is due to various current management mechanisms that throttle SIMD instruction throughput while the processor adjusts voltage and frequency settings~\cite{haj2021ichannels}.
Padded \AttackName can detect this penalty in AVX-512 instructions.
For this experiment, we replace the memory access to the probe address in \tprobe with an AVX-512 fused multiply-add instruction.
Specifically, we add a single \texttt{vfnmsub231pd} instruction with two \texttt{vmov} instructions to transfer input and output between SIMD and GP registers.
We also remove the reference memory access in \tref, relying entirely on the padding in \tprobe to separate the two cases.
For the fast case, we zero all AVX registers and then pre-warm the AVX-512 units by spinning on a series of AVX-512 instructions for \SI{5}{\micro\second} before measurement.
For the slow case, we spin on \texttt{rdtsc} for \SI{1}{\milli\second}, not executing any AVX instructions.
We run this experiment on an Intel \labeightynineCPU with the threads pinned to cores 0 and 1 (i.e., separate physical cores), taking \num{512} measurement samples per case and padding level.
Starting at a padding level of \num{159} \texttt{add}-cycles, we observe a golden range of \num{222} \texttt{add}-cycles in which the combined $F_1$ score is above \num{0.98}.
In all but \num{5} outlier cases in this range, the $F_1$ score is $> 0.995$.
This shows that \AttackName can reliably detect this timing difference.

\paragrabf{FP Operations with Subnormal Operands}
Floating-point standards such as IEEE-754~\cite{ieee7542008} define a special number class known as subnormal (or sometimes called denormal) numbers.
In the binary interchange formats, a subnormal is encoded with an all-zero biased exponent field and a non-zero trailing significand, allowing the representation of values very close to zero.
This fills an otherwise unused part of the encoding space but adds complexity when implementing IEEE-754-compliant floating-point arithmetic in hardware.
Operations involving subnormal numbers often incur a performance penalty~\cite{Andrysco2015subnormal,kohlbrenner2017effectiveness}, with Intel CPUs typically relying on microcode assists for their implementation~\cite{Intel2023optimization}.
To measure this with \AttackName, we replace the memory access to the probe address in \tprobe with an AVX2 floating-point division, and let it operate on normal numbers for the fast case and subnormal numbers for the slow case.
As in the previous experiment, we remove the reference memory access in \tref.
We run this experiment on SMT sibling cores of an Intel \labeightynineCPU, taking \num{4096} samples per configuration and padding level.
We observe an $F_1$ score of \num{0.9991} at padding level $0$, with the golden range extending to \num{31} \texttt{add}-cycles.
Hence, this timing difference can be reliably detected even with unpadded \AttackName.

\paragrabf{TLB Residency}
\AttackName can detect whether a CPU's TLB caches a translation for a given virtual address.
We verify this with an experiment on an Intel \labfourtytwoCPU.
In the fast case, we access the probe address before measurement.
In the slow case, we clear it from the TLB using the \texttt{invlpg} instruction and access a virtual alias of the probe address's physical location.
This ensures that any timing differences originate only from the TLB, with the address in the L1d cache in both cases.
For each case, we take \num{16384} samples per padding level with the threads pinned to separate physical cores.
We observe a golden range of \num{39} \texttt{add}-cycles, starting at padding level \num{139}, meaning the padded \AttackName can reliably detect the timing difference.
If we additionally flush the relevant page table entries to DRAM with \texttt{clflush} before probing, we observe perfect accuracy even without padding, and a golden range of \num{178} \texttt{add}-cycles.

\subsection{\AttackName in Lua}\label{subsec:attackname-in-lua}

While previous experiments rely on direct access to binary code or simple environments like WebAssembly to implement \AttackName, the low-level capabilities these provide are not strictly required.
To demonstrate that \AttackName remains effective in a higher-level language, we devise an experiment with unpadded \AttackName for the LuaJIT environment (version 2.1.1703358377).
It follows the same structure as the WebAssembly implementation in \Cref{subsec:attackname-in-the-browser}, with the primitive implemented entirely in Lua and a simple C library to prepare the target address's cache state.
Additionally, we use LuaJIT's foreign function interface to create a shared memory region.
As with WebAssembly, we omit cache state preparation for $w$ and $s$.
We evaluate this experiment in the L1 vs.\ DRAM configuration, running on an AMD \labthirtyCPU with the threads on separate physical cores.
Overall, we take \num{1000000} samples each for the cached and the uncached case.
We observe just \num{120} misclassifications in the cached case and \num{150} in the uncached case, resulting in a combined $F_1$ score of $> 0.999$.
This shows that \AttackName is feasible in higher-level languages.

\section{Discussion}\label{sec:discussion}
In this section, we discuss \AttackName's limitations, potential mitigations, and related work.

\subsection{Limitations}\label{subsec:limitations}

\paragrabf{No Direct Timing Measurement}
Compared to a precise hardware timer, \AttackName's primary limitation is that it does not measure timing directly.
Instead, it compares two timings and reveals which is faster.
In cache attacks, this is typically not an issue.
Attacks like Flush+Reload or Prime+Probe reveal only whether an address was accessed, so absolute timing values are irrelevant.
However, this complicates attacks with more than two states to distinguish (cf. \Cref{subsec:determining-llc-slices}, where there are 12).
In this scenario, attacks using \AttackName need multiple measurements, whereas a precise timer can, in theory, distinguish them with a single sample.

\paragrabf{Precisely Controlled Padding}
The padded variant requires a mechanism for selecting a precisely controlled delay sequence.
Our native and WebAssembly implementations generate the required dependency chain at runtime.
Many higher-level languages do not expose comparable control over generated code.
Such environments can instead use loop-based padding or ship a finite collection of precompiled padding functions.
Both approaches may introduce additional jitter or provide coarser calibration steps.

\subsection{Related Work}\label{subsec:related-work}

\paragrabf{Timerless Cache Side Channels}
In environments without a timer, a standard approach for measuring timing differences is to create a timer using counting threads~\cite{Lipp2016,Schwarz2017Timers,Schwarz2020malware}.
We compare this approach to \AttackName in \Cref{subsec:comparison-to-counter-threads}.
The idea of using race conditions with thread-level parallelism for revealing memory access latencies, and thus the foundational principle of \AttackName, was proposed by Purnal \etal under the name \emph{architectural reordering}~\cite{Purnal2023showtime}.
Our work extensively evaluates and extends this idea.
Other fully timerless cache side channels typically rely on architecture-specific properties.
For instance, the abort semantics of Intel TSX may reveal cache state information~\cite{Disselkoen2017prime,kim2022dprime,Hornetz2026tdxray}.
Similarly, weak coherence between the data and instruction caches on many RISC architectures is known to enable timerless attack primitives~\cite{Guanciale2016Cache,Jin2025Ghostcache,Thomas2025Exfilstate,thomas2026riscy}.
Zhang \etal show that the \texttt{umonitor} and \texttt{umwait} instructions on x86-based architectures provide unprivileged users with architectural feedback on cache activity~\cite{Zhang2023MWAIT}.
Yu \etal create S${}^2$C, a timerless cache side channel for Apple's M1 architecture based on hardware synchronization instructions~\cite{Yu2023Synchronization}.
Thomas \etal propose ExfilState, a fuzzing-based technique for discovering architectural cache side channels on ARM~\cite{Thomas2025Exfilstate}.
In a follow-up work, Thomas~\etal\cite{thomas2026riscy} show that cache coherency effects in RISC CPUs can reveal information on the instruction cache's state.

\paragrabf{Attacks with Imprecise Timers}
In addition to fully timerless attacks like \AttackName, a large body of work exists on cache attacks with imprecise timers.
This line of work is orthogonal to \AttackName, as the limiting factor in cache attacks is often the resolution of a timer, not its full absence.
Many rely on \emph{amplification}, i.e., magnifying the timing differences between the cached and uncached states until they are measurable even with a coarse-grained timer.
A seminal work is \texttt{leaky.page}~\cite{Roettger2021SpectreJS}, which introduces an amplification primitive that relies on the Tree-PLRU cache replacement policy used in the L1 caches of many microarchitectures.
Purnal \etal extend this primitive, significantly increasing its amplification ratio~\cite{Purnal2023showtime}, and propose another amplification primitive using non-temporal prefetch instructions on x86.
Katzman \etal propose an amplification method using logic gates operating on the cache state~\cite{Katzman2023gates}.
It propagates the cache state of a target address in a tree-like structure, creating an amplified timing difference when accessing the leaves sequentially.
Morgan \etal introduce comparator gates, which create a race condition between two memory accesses during speculative execution, encoding the result in the cache~\cite{Morgan2025slice}.
This amplifies a timing difference on the order of individual cycles to a difference between L1 and DRAM\@.
Xiao \etal propose Hacky Racers~\cite{xiao2023hacky}, a primitive relying on instruction-level parallelism introduced by out-of-order execution.
It encodes the relative completion order of two parallelizable instruction paths in the cache and uses amplification to make it accessible with a coarse-grained timer.
An orthogonal approach to amplification is to average many direct measurements with a coarse-grained timer.
This requires many samples per target and was demonstrated mainly in remote attacks~\cite{Bernstein2005,Schwarz2019netspectre,Schwarzl2021dynamic}.

\subsection{Mitigations}\label{subsec:mitigations}

\paragrabf{Eliminating Shared Memory}
Shared memory, a requirement for \AttackName, can often be replaced with message passing via channels.
This alone does not fully mitigate \AttackName, as it could, in theory, also be implemented with channels by replacing writes to $s$ and $w$ with channel writes.
However, channels are often managed components, e.g., by the operating system, and thus provide a surface for implementing mitigations.
In many web browsers, for instance, message passing is already rate-limited to prevent the construction of counting threads~\cite{Kohlbrenner2016,Schwarz2017Timers}.
To further mitigate \AttackName, channels can add a random delay before delivering a message.
This makes the race-condition outcome random, preventing attackers from inferring usable information from a single measurement.
Another mitigation is to grant write access to a shared memory segment to only one thread at a time.
\AttackName requires $w$ to be writable by both threads, which such a configuration prevents.

\paragrabf{Eliminating True Parallelism}
Many programs follow an asynchronous programming model, forgoing true parallelism in favor of weaker, interleaving-based concurrency.
While this does not bring the full performance benefits of parallelism, it can unblock expensive operations, allowing programs to continue executing while waiting (e.g., for the operating system).
A prominent example is asynchronous functions in languages such as JavaScript.
Implementing \AttackName is not possible in such an environment, as it requires parallelism.

\section{Conclusion}\label{sec:conclusion}
We presented \AttackName, a timerless probing primitive for distinguishing microarchitectural latencies.
In contrast to other timerless primitives, \AttackName is portable across microarchitectures and relies only on thread-level parallelism and shared memory.
We performed an extensive evaluation of \AttackName on \PlatformCount CPUs, demonstrating that it can reliably separate microarchitectural latency differences on all of them.
Additionally, we showed that \AttackName works in a web browser, with a temporal accuracy approaching that of our native implementations.
Our case studies showed that \AttackName applies in a wide variety of contexts and can, in some instances, distinguish latency differences on the order of individual cycles.
Overall, \AttackName shows that timing-based microarchitectural attacks, even with fine-grained latency differences, do not inherently require a timer.

\FloatBarrier

\bibliographystyle{plain}
\bibliography{main}
\newpage

\appendix
\crefalias{section}{appendix}
\FloatBarrier

\section{Per-Machine Experimental Results}\label{sec:precision-analysis}
We present the per-machine experimental results from \Cref{subsec:delta-sync} and \Cref{subsec:measuring-memory-access-latencies} in \Cref{tab:precision} and \Cref{tab:precision_ht}.
$F^{fast}_{slow}$ denotes the $F_1$ score of the unpadded \AttackName in a specific setting, and $G^{fast}_{slow}$ denotes the width of the golden range with padded \AttackName in \texttt{add}-cycles.
$CTR$ is the average edge width of a counting thread, and $\Delta^{L1}_{DRAM}$ is the latency difference between an L1 hit and a DRAM access as measured in \Cref{subsec:delta-sync}.
\Cref{tab:precision_wasm} lists the results of our WASM-based implementation on x86\_64.

\section{Reliable Cache Eviction}\label{sec:reliable-cache-eviction}
Many experiments in this paper rely on eviction sets to prime the cache state, e.g., to evict an address from the L2 or L3 cache.
Unless otherwise specified, as in \Cref{subsec:timerless-evict-reload}, we construct these eviction sets using the physical addresses of both the target and eviction-set addresses.
Our eviction sets consist of $n + 2$ set-index-congruent addresses for evicting the L1 cache, and $n + 4$ addresses for evicting the L2 cache, where $n$ is the associativity of the targeted cache level.
The overfill provides additional robustness.
On CPUs where the L1 cache has higher associativity than the L2, such as Intel's Skylake architectures, we use the larger L1 associativity for $n$ when evicting the L2 cache.
To apply an eviction set, we first access the target address to cache it in the L1 cache.
Then, we access all eviction-set addresses in a linear pass, then in multiple passes in random order, and then again in a reverse-order linear pass.
To our knowledge, all L1 caches in our evaluation are directly indexed, i.e., there is no complex hash function for determining the cache set index.
Testing for congruency is therefore trivial.
With the L2 cache, we assume direct indexing for all CPUs except for Apple's M1 and M2 architectures and AMD's Zen 3, 4, and 5 architectures.
For M1, we rely on the set index function reverse-engineered by Yu et al.~\cite{Yu2023Synchronization}, and for M2, we find L2 eviction sets using the algorithm described by Vila et al.~\cite{Vila2019}.
For AMD, we use the set index functions reverse-engineered by Bagia et al.~\cite{bagia2025close} and Zhang et al.~\cite{Zhang2025Taming} to determine cache set congruency.
Where possible, we rely on cache management instructions to flush addresses from the LLC\@.
On architectures where such instructions are privileged, we use a simple kernel module to grant user-mode processes access.
We still perform LLC eviction with eviction sets only on the StarFive JH7100 and the Microchip MPFS025T.
Based on publicly available RTL code~\cite{sifive-inclusivecache}, we assume direct set indexing and construct the LLC eviction sets in the same way as L2 eviction sets on other platforms.

\begin{table*}[b]
  \caption{\AttackName characteristics with \tprobe on core 0 and \tref on the SMT sibling core. We list \rdelta{sync}, $\Delta^{L1}_{DRAM}$, and the average edge width of a counting thread (CTR). Additionally, it shows the $F_1$ score of unpadded race probing ($F^{fast}_{slow}$) and the golden range's length for padded race probing ($G^{fast}_{slow}$). All values, except the $F_1$ scores, are measured in \texttt{add}-cycles.}
  \label{tab:precision_ht}
  \centering
  \footnotesize
\renewcommand{\arraystretch}{0.75}
\begin{tabular}{cll@{\hspace{1em}}rrr@{\hspace{2.5em}}cc@{\hspace{2.5em}}cccc}
\toprule & \multirow[c]{2}{*}[-3pt]{\textbf{CPU}} & \multirow[c]{2}{*}[-3pt]{\textbf{$\mu$Architecture}} &&&& \multicolumn{2}{c}{\hspace*{-2.5em}\textbf{Unpadded}} & \multicolumn{4}{c}{\hspace*{-\tabcolsep}\textbf{Padded}} \vspace{3.5pt}\\
& &  & $\Delta_{sync}$ & $\Delta^{L1}_{DRAM}$ & CTR & $F^{L1}_{DRAM}$ & $F^{LLC}_{DRAM}$ & $G^{L1}_{DRAM}$ & $G^{L1}_{L2}$ & $G^{L1}_{L3}$ & $G^{LLC}_{DRAM}$ \vspace{1pt}\\\midrule
\multirow[c]{19}{*}{\rotatebox[origin=c]{90}{\textbf{x86}\hspace{11em}}} & AMD EPYC 7252 & Zen 2 (Rome) & $43 \pm 1$ & 329 & 74 & 1.00 & 0.99 & 41 & 5 & 31 & 12 \\\cmidrule(lr){2-12}
 & AMD Ryzen 7 5700G & Zen 3 (Cezanne) & $55 \pm 22$ & 322 & 90 & 0.99 & 0.77 & 6 & $\cdot$ & $\cdot$ & $\cdot$ \\\cmidrule(lr){2-12}
 & AMD Ryzen 9 7950X3D & Zen 4 (Raphael) & $38 \pm 29$ & 496 & 95 & 0.99 & 0.61 & 5 & $\cdot$ & $\cdot$ & $\cdot$ \\\cmidrule(lr){2-12}
 & AMD EPYC 9124 & Zen 4 (Genoa) & $50 \pm 19$ & 324 & 96 & 0.99 & 0.23 & 9 & $\cdot$ & 2 & $\cdot$ \\\cmidrule(lr){2-12}
 & AMD Ryzen 9 9950X & Zen 5 (Granite Ridge) & $76 \pm 31$ & 462 & 249 & 0.99 & 0.93 & 37 & $\cdot$ & $\cdot$ & $\cdot$ \\\cmidrule(lr){2-12}
 & Intel Core i3-2120 & Sandy Bridge & $47 \pm 8$ & 200 & 50 & 1.00 & 1.00 & 36 & $\cdot$ & $\cdot$ & 43 \\\cmidrule(lr){2-12}
 & Intel Core i3-5010U & Broadwell & $76 \pm 7$ & 156 & 43 & 0.99 & 0.99 & 18 & $\cdot$ & $\cdot$ & 14 \\\cmidrule(lr){2-12}
 & Intel Xeon E3-1505M v5 & Skylake & $35 \pm 13$ & 170 & 42 & 1.00 & 0.81 & 20 & $\cdot$ & $\cdot$ & $\cdot$ \\\cmidrule(lr){2-12}
 & Intel Pentium Gold G5400T & Coffee Lake & $44 \pm 8$ & 205 & 41 & 1.00 & 0.94 & 39 & $\cdot$ & 25 & $\cdot$ \\\cmidrule(lr){2-12}
 & Intel Core i7-8700 & Coffee Lake & $35 \pm 18$ & 272 & 25 & 1.00 & 0.24 & 39 & $\cdot$ & 39 & $\cdot$ \\\cmidrule(lr){2-12}
 & Intel Core i7-10710U & Comet Lake & $35 \pm 15$ & 305 & 43 & 1.00 & 0.92 & 20 & $\cdot$ & $\cdot$ & $\cdot$ \\\cmidrule(lr){2-12}
 & Intel Core i7-1185G7 & Willow Cove & $54 \pm 21$ & 513 & 41 & 0.99 & 0.88 & 32 & $\cdot$ & 8 & $\cdot$ \\\cmidrule(lr){2-12}
 & Intel Xeon Gold 6346 & Sunny Cove & $47 \pm 25$ & 233 & 47 & 1.00 & 1.00 & 27 & $\cdot$ & 26 & 31 \\\cmidrule(lr){2-12}
 & Intel Core i7-11700 & Cypress Cove & $53 \pm 20$ & 316 & 46 & 0.99 & 0.98 & 28 & $\cdot$ & 18 & 6 \\\cmidrule(lr){2-12}
 & Intel Core i9-12900H & Golden Cove (P) & $68 \pm 22$ & 345 & 49 & 1.00 & 0.99 & 40 & $\cdot$ & 14 & 53 \\\cmidrule(lr){2-12}
 & Intel Core i5-13420H & Raptor Cove (P) & $71 \pm 18$ & 345 & 56 & 1.00 & 0.99 & 45 & $\cdot$ & 18 & 50 \\\cmidrule(lr){2-12}
 & Intel Core i9-13900K & Raptor Cove (P) & $70 \pm 15$ & 456 & 58 & 0.99 & 0.94 & 43 & $\cdot$ & 25 & $\cdot$ \\\cmidrule(lr){2-12}
 & Intel Xeon Gold 6526Y & Raptor Cove & $54 \pm 30$ & 357 & 58 & 1.00 & 0.10 & 42 & $\cdot$ & 35 & $\cdot$ \\\cmidrule(lr){2-12}
 & Intel Xeon 6736P & Redwood Cove & $60 \pm 17$ & 681 & 50 & 1.00 & 0.94 & 45 & $\cdot$ & 39 & $\cdot$ \\\midrule
\multirow[c]{1}{*}{\textbf{LA64}} & Loongson 3A6000 & 3A6000 & $22 \pm 8$ & 183 & 3 & 1.00 & 0.00 & 14 & $\cdot$ & 14 & $\cdot$ \\
\bottomrule
\end{tabular}

\end{table*}
\begin{table*}[t]
  \caption{\AttackName characteristics with \tprobe on core 0 and \tref on the physical core with the next-highest ID. We list \rdelta{sync}, $\Delta^{L1}_{DRAM}$, and the average edge width of a counting thread (CTR). Additionally, it shows the $F_1$ score of unpadded race probing ($F^{fast}_{slow}$) and the golden range's length for padded race probing ($G^{fast}_{slow}$). All values, except the $F_1$ scores, are measured in \texttt{add}-cycles.}
  \label{tab:precision}
  \centering
  \footnotesize
\renewcommand{\arraystretch}{0.75}
\begin{tabular}{cll@{\hspace{1em}}rrr@{\hspace{2.5em}}cc@{\hspace{2.5em}}cccc}
\toprule & \multirow[c]{2}{*}[-3pt]{\textbf{CPU}} & \multirow[c]{2}{*}[-3pt]{\textbf{$\mu$Architecture}} &&&& \multicolumn{2}{c}{\hspace*{-2.5em}\textbf{Unpadded}} & \multicolumn{4}{c}{\hspace*{-\tabcolsep}\textbf{Padded}} \vspace{3.5pt}\\
& &  & $\Delta_{sync}$ & $\Delta^{L1}_{DRAM}$ & CTR & $F^{L1}_{DRAM}$ & $F^{LLC}_{DRAM}$ & $G^{L1}_{DRAM}$ & $G^{L1}_{L2}$ & $G^{L1}_{L3}$ & $G^{LLC}_{DRAM}$ \vspace{1pt}\\\midrule
\multirow[c]{8}{*}{\rotatebox[origin=c]{90}{\textbf{ARM}\hspace{4em}}} & \multirow[c]{2}{*}[-2.25pt]{Apple M1} & Icestorm (E) & $122 \pm 1$ & 236 & 86 & 0.99 & 0.99 & 177 & 8 & 88 & 87 \\\cmidrule(lr){3-12}
 &  & Firestorm (P) & $142 \pm 1$ & 342 & 105 & 1.00 & 0.99 & 230 & 53 & 122 & 44 \\\cmidrule(lr){2-12}
 & \multirow[c]{2}{*}[-2.25pt]{Apple M2} & Blizzard (E) & $123 \pm 6$ & 290 & 97 & 0.97 & 0.99 & 123 & $\cdot$ & 24 & 65 \\\cmidrule(lr){3-12}
 &  & Avalanche (P) & $142 \pm 7$ & 371 & 112 & 0.97 & 0.99 & 92 & $\cdot$ & $\cdot$ & 355 \\\cmidrule(lr){2-12}
 & Broadcom BCM2837 & Cortex-A53 & $49 \pm 0$ & 161 & 35 & 0.99 & 0.99 & 50 & 13 & -- & 37 \\\cmidrule(lr){2-12}
 & Broadcom BCM2712 & Cortex-A76 & $143 \pm 61$ & 322 & 172 & 0.99 & 0.99 & 161 & $\cdot$ & $\cdot$ & 234 \\\cmidrule(lr){2-12}
 & Huawei Kunpeng 920 & TaiShan-v110 & $243 \pm 1$ & 271 & 91 & 0.75 & 0.75 & 104 & $\cdot$ & $\cdot$ & 92 \\\cmidrule(lr){2-12}
 & Ampere Altra Q64-30 & Altra & $293 \pm 20$ & 251 & 242 & 0.67 & 0.67 & 99 & $\cdot$ & $\cdot$ & 120 \\\midrule
\multirow[c]{31}{*}{\rotatebox[origin=c]{90}{\textbf{x86}\hspace{18em}}} & AMD EPYC 7252 & Zen 2 (Rome) & $78 \pm 0$ & 329 & 205 & 0.99 & 0.99 & 78 & 10 & 34 & 44 \\\cmidrule(lr){2-12}
 & AMD Ryzen 7 5700G & Zen 3 (Cezanne) & $84 \pm 0$ & 322 & 162 & 0.99 & 0.99 & 71 & $\cdot$ & 33 & 37 \\\cmidrule(lr){2-12}
 & AMD Ryzen 9 7950X3D & Zen 4 (Raphael) & $87 \pm 3$ & 496 & 183 & 1.00 & 0.99 & 66 & $\cdot$ & 6 & 38 \\\cmidrule(lr){2-12}
 & AMD EPYC 9124 & Zen 4 (Genoa) & $85 \pm 5$ & 324 & 195 & 0.99 & 0.84 & 49 & $\cdot$ & $\cdot$ & $\cdot$ \\\cmidrule(lr){2-12}
 & AMD Ryzen 9 9950X & Zen 5 (Granite Ridge) & $140 \pm 1$ & 462 & 201 & 0.99 & 0.99 & 125 & 10 & 47 & 78 \\\cmidrule(lr){2-12}
 & Intel Core i3-2120 & Sandy Bridge & $71 \pm 8$ & 200 & 102 & 0.99 & 0.99 & 61 & 2 & 7 & 62 \\\cmidrule(lr){2-12}
 & Intel Core i5-4690 & Haswell & $95 \pm 1$ & 281 & 116 & 0.99 & 1.00 & 89 & $\cdot$ & 13 & 100 \\\cmidrule(lr){2-12}
 & Intel Core i3-5010U & Broadwell & $88 \pm 3$ & 156 & 117 & 0.99 & 0.99 & 88 & 2 & 12 & 54 \\\cmidrule(lr){2-12}
 & Intel Xeon E3-1505M v5 & Skylake & $90 \pm 1$ & 170 & 126 & 1.00 & 1.00 & 97 & 2 & 21 & 68 \\\cmidrule(lr){2-12}
 & Intel Core i5-6400T & Skylake & $95 \pm 5$ & 88 & 134 & 0.68 & 0.68 & 80 & $\cdot$ & 19 & 50 \\\cmidrule(lr){2-12}
 & Intel Pentium G4400T & Skylake & $88 \pm 4$ & 190 & 120 & 0.99 & 0.99 & 92 & $\cdot$ & $\cdot$ & 63 \\\cmidrule(lr){2-12}
 & Intel Pentium Gold G5400T & Coffee Lake & $92 \pm 3$ & 205 & 146 & 1.00 & 1.00 & 91 & $\cdot$ & 22 & 59 \\\cmidrule(lr){2-12}
 & Intel Core i7-8700 & Coffee Lake & $96 \pm 3$ & 272 & 114 & 1.00 & 0.99 & 95 & $\cdot$ & 27 & 54 \\\cmidrule(lr){2-12}
 & Intel Core i7-9700K & Coffee Lake & $91 \pm 5$ & 298 & 156 & 1.00 & 0.99 & 103 & $\cdot$ & 22 & 73 \\\cmidrule(lr){2-12}
 & Intel Core i7-10710U & Comet Lake & $95 \pm 3$ & 305 & 224 & 0.99 & 0.99 & 90 & $\cdot$ & 21 & 57 \\\cmidrule(lr){2-12}
 & Intel Core i7-1185G7 & Willow Cove & $126 \pm 17$ & 513 & 231 & 0.99 & 0.99 & 103 & $\cdot$ & 9 & 55 \\\cmidrule(lr){2-12}
 & Intel Xeon Gold 6346 & Sunny Cove & $151 \pm 6$ & 233 & 300 & 1.00 & 0.69 & 169 & 12 & 50 & 143 \\\cmidrule(lr){2-12}
 & Intel Core i7-11700 & Cypress Cove & $134 \pm 12$ & 316 & 241 & 0.99 & 1.00 & 131 & $\cdot$ & $\cdot$ & 83 \\\cmidrule(lr){2-12}
 & \multirow[c]{2}{*}[-2.25pt]{Intel Core i9-12900H} & Gracemont (E) & $147 \pm 3$ & 295 & 194 & 1.00 & 1.00 & 147 & 14 & 68 & 74 \\\cmidrule(lr){3-12}
 &  & Golden Cove (P) & $145 \pm 22$ & 345 & 244 & 0.99 & 0.99 & 133 & $\cdot$ & 24 & 82 \\\cmidrule(lr){2-12}
 & \multirow[c]{2}{*}[-2.25pt]{Intel Core i5-13420H} & Gracemont (E) & $156 \pm 4$ & 288 & 198 & 0.99 & 0.99 & 157 & 7 & 33 & 113 \\\cmidrule(lr){3-12}
 &  & Raptor Cove (P) & $142 \pm 15$ & 345 & 258 & 0.99 & 0.99 & 92 & $\cdot$ & 30 & 21 \\\cmidrule(lr){2-12}
 & \multirow[c]{2}{*}[-2.25pt]{Intel Core i9-13900K} & Gracemont (E) & $170 \pm 10$ & 392 & 166 & 1.00 & 1.00 & 167 & 14 & 52 & 97 \\\cmidrule(lr){3-12}
 &  & Raptor Cove (P) & $179 \pm 13$ & 456 & 357 & 0.99 & 0.99 & 156 & $\cdot$ & 31 & 61 \\\cmidrule(lr){2-12}
 & Intel Xeon Gold 6526Y & Raptor Cove & $143 \pm 8$ & 357 & 355 & 0.99 & 0.99 & 169 & 8 & 65 & 93 \\\cmidrule(lr){2-12}
 & Intel Core Ultra 9 288V & Lion Cove & $153 \pm 13$ & 279 & 286 & 0.99 & 0.99 & 135 & $\cdot$ & 38 & 71 \\\cmidrule(lr){2-12}
 & \multirow[c]{2}{*}[-2.25pt]{Intel Core Ultra 9 285K} & Skymont (E) & $98 \pm 0$ & 475 & 86 & 1.00 & 1.00 & 97 & $\cdot$ & 10 & 33 \\\cmidrule(lr){3-12}
 &  & Lion Cove (P) & $162 \pm 17$ & 577 & 266 & 0.99 & 0.99 & 234 & $\cdot$ & 40 & 172 \\\cmidrule(lr){2-12}
 & Intel Xeon 6736P & Redwood Cove & $403 \pm 23$ & 681 & 792 & 0.99 & 0.99 & 396 & $\cdot$ & 138 & 193 \\\cmidrule(lr){2-12}
 & \multirow[c]{2}{*}[-2.25pt]{Intel Core Ultra 5 325} & Darkmont (E) & $90 \pm 7$ & 121 & 86 & 0.94 & 0.93 & 63 & $\cdot$ & $\cdot$ & 47 \\\cmidrule(lr){3-12}
 &  & Cougar Cove (P) & $151 \pm 25$ & 284 & 202 & 1.00 & 0.99 & 150 & $\cdot$ & 28 & 88 \\\midrule
\multirow[c]{4}{*}{\rotatebox[origin=c]{90}{\textbf{RISC-V}\hspace{1em}}} & StarFive JH7100 & U74 & $27 \pm 15$ & 129 & 55 & 0.98 & 0.59 & 2 & $\cdot$ & -- & $\cdot$ \\\cmidrule(lr){2-12}
 & Microchip MPFS025T & U54 & $18 \pm 0$ & 75 & 54 & 0.99 & 0.66 & 19 & $\cdot$ & -- & $\cdot$ \\\cmidrule(lr){2-12}
 & T-Head TH1520 & C910 & $56 \pm 0$ & 239 & 86 & 0.99 & 0.99 & 48 & 29 & -- & 48 \\\cmidrule(lr){2-12}
 & SpacemiT K1 & X60 & $75 \pm 7$ & 145 & 216 & 0.99 & 1.00 & 55 & 2 & -- & 45 \\\midrule
\multirow[c]{2}{*}{\rotatebox[origin=c]{90}{\hspace{.5em}\textbf{LA64}}} & Loongson 3A5000 & 3A5000 & $186 \pm 2$ & 309 & 123 & 0.99 & 1.00 & 150 & $\cdot$ & 5 & 140 \\\cmidrule(lr){2-12}
 & Loongson 3A6000 & 3A6000 & $120 \pm 2$ & 183 & 100 & 1.00 & 1.00 & 119 & $\cdot$ & 33 & 84 \\
\bottomrule
\end{tabular}

\end{table*}
\begin{table*}[t]
    \caption{\AttackName performance in WebAssembly (Chromium v147.0) in the L1 vs. DRAM setting on x86\_64. \rdelta{sync} and the width of the golden range $G^{L1}_{DRAM}$ are measured in \texttt{add}-cycles.}
    \label{tab:precision_wasm}
    \centering
    \footnotesize
\renewcommand{\arraystretch}{0.75}
\begin{tabular}{ll@{\hspace{2em}}rr@{\hspace{3em}}cc@{\hspace{3em}}cc}
\toprule \multirow[c]{2}{*}[-3pt]{\textbf{CPU}} & \multirow[c]{2}{*}[-3pt]{\textbf{$\mu$Architecture}} & \multicolumn{2}{c}{\hspace*{-3em}\makebox[0pt]{$\Delta_{sync}$}} & \multicolumn{2}{c}{\hspace*{-3em}\makebox[0pt]{$F^{L1}_{DRAM}$ (unpadded)}} & \multicolumn{2}{c}{\hspace*{-\tabcolsep}\makebox[0pt]{$G^{L1}_{DRAM}$ (padded)}} \vspace{1pt}\\\cmidrule(lr{\dimexpr 3em+0.5em\relax}){3-4}\cmidrule(lr{\dimexpr 3em+0.5em\relax}){5-6}\cmidrule(lr{\dimexpr \tabcolsep+0.5em\relax}){7-8}
 & & SMT & Cross-core & SMT & Cross-core & SMT & Cross-core \vspace{1pt}\\\midrule
AMD EPYC 7252 & Zen 2 (Rome) & $50 \pm 4$ & $101 \pm 33$ & 0.99 & 0.99 & 38 & 67 \\\cmidrule(lr){1-8}
AMD Ryzen 7 5700G & Zen 3 (Cezanne) & $48 \pm 20$ & $126 \pm 3$ & 0.99 & 0.99 & 44 & 113 \\\cmidrule(lr){1-8}
AMD Ryzen 9 7950X3D & Zen 4 (Raphael) & $49 \pm 21$ & $88 \pm 13$ & 0.99 & 0.99 & 50 & 69 \\\cmidrule(lr){1-8}
AMD EPYC 9124 & Zen 4 (Genoa) & $48 \pm 18$ & $120 \pm 1$ & 0.99 & 0.99 & 42 & 128 \\\cmidrule(lr){1-8}
AMD Ryzen 9 9950X & Zen 5 (Granite Ridge) & $91 \pm 7$ & $165 \pm 7$ & 0.99 & 0.99 & 69 & 137 \\\cmidrule(lr){1-8}
Intel Core i3-2120 & Sandy Bridge & $24 \pm 12$ & $84 \pm 30$ & 0.99 & 0.98 & 22 & 36 \\\cmidrule(lr){1-8}
Intel Core i5-4690 & Haswell & -- & $83 \pm 1$ & -- & 0.99 & -- & 69 \\\cmidrule(lr){1-8}
Intel Core i3-5010U & Broadwell & $28 \pm 14$ & $108 \pm 4$ & 0.99 & 0.99 & 25 & 61 \\\cmidrule(lr){1-8}
Intel Xeon E3-1505M v5 & Skylake & $33 \pm 20$ & $89 \pm 7$ & 0.99 & 0.99 & 41 & 66 \\\cmidrule(lr){1-8}
Intel Core i5-6400T & Skylake & -- & $99 \pm 5$ & -- & 0.97 & -- & 50 \\\cmidrule(lr){1-8}
Intel Pentium G4400T & Skylake & -- & $91 \pm 43$ & -- & 0.99 & -- & 26 \\\cmidrule(lr){1-8}
Intel Pentium Gold G5400T & Coffee Lake & $36 \pm 17$ & $128 \pm 6$ & 0.99 & 0.99 & 30 & 98 \\\cmidrule(lr){1-8}
Intel Core i7-8700 & Coffee Lake & $36 \pm 17$ & $110 \pm 26$ & 0.99 & 0.99 & 28 & 75 \\\cmidrule(lr){1-8}
Intel Core i7-9700K & Coffee Lake & -- & $106 \pm 4$ & -- & 0.98 & -- & 82 \\\cmidrule(lr){1-8}
Intel Core i7-10710U & Comet Lake & $32 \pm 19$ & $120 \pm 12$ & 0.99 & 0.99 & 11 & 29 \\\cmidrule(lr){1-8}
Intel Core i7-1185G7 & Willow Cove & $46 \pm 11$ & $122 \pm 18$ & 0.99 & 0.99 & 41 & 90 \\\cmidrule(lr){1-8}
Intel Xeon Gold 6346 & Sunny Cove & $36 \pm 17$ & $327 \pm 113$ & 0.99 & 0.99 & 37 & 316 \\\cmidrule(lr){1-8}
Intel Core i7-11700 & Cypress Cove & $39 \pm 6$ & $130 \pm 5$ & 0.99 & 0.99 & 11 & 88 \\\cmidrule(lr){1-8}
\multirow[c]{2}{*}[-2.25pt]{Intel Core i9-12900H} & Gracemont (E) & -- & $140 \pm 5$ & -- & 0.99 & -- & 125 \\\cmidrule(lr){2-8}
 & Golden Cove (P) & $55 \pm 16$ & $229 \pm 37$ & 0.99 & 0.99 & 27 & 139 \\\cmidrule(lr){1-8}
\multirow[c]{2}{*}[-2.25pt]{Intel Core i5-13420H} & Gracemont (E) & -- & $128 \pm 6$ & -- & 0.99 & -- & 73 \\\cmidrule(lr){2-8}
 & Raptor Cove (P) & $55 \pm 16$ & $193 \pm 17$ & 0.99 & 0.99 & 17 & 155 \\\cmidrule(lr){1-8}
\multirow[c]{2}{*}[-2.25pt]{Intel Core i9-13900K} & Gracemont (E) & -- & $168 \pm 4$ & -- & 0.99 & -- & 123 \\\cmidrule(lr){2-8}
 & Raptor Cove (P) & $54 \pm 15$ & $276 \pm 16$ & 0.99 & 0.99 & 31 & 188 \\\cmidrule(lr){1-8}
Intel Xeon Gold 6526Y & Raptor Cove & $49 \pm 28$ & $303 \pm 21$ & 0.99 & 0.99 & 33 & 191 \\\cmidrule(lr){1-8}
Intel Core Ultra 9 288V & Lion Cove & -- & $173 \pm 23$ & -- & 0.99 & -- & 152 \\\cmidrule(lr){1-8}
\multirow[c]{2}{*}[-2.25pt]{Intel Core Ultra 9 285K} & Skymont (E) & -- & $180 \pm 2$ & -- & 0.97 & -- & 142 \\\cmidrule(lr){2-8}
 & Lion Cove (P) & -- & $326 \pm 31$ & -- & 0.99 & -- & 218 \\\cmidrule(lr){1-8}
Intel Xeon 6736P & Redwood Cove & $53 \pm 20$ & $974 \pm 31$ & 0.99 & 0.99 & 35 & 361 \\\cmidrule(lr){1-8}
\multirow[c]{2}{*}[-2.25pt]{Intel Core Ultra 5 325} & Darkmont (E) & -- & $130 \pm 100$ & -- & 0.99 & -- & 14 \\\cmidrule(lr){2-8}
 & Cougar Cove (P) & -- & $172 \pm 12$ & -- & 0.99 & -- & 128 \\
\bottomrule
\end{tabular}

\end{table*}

\end{document}